\documentclass[12pt]{article}

\usepackage[letterpaper,margin=1in]{geometry}
\usepackage{amsmath,amssymb}
\usepackage{xcolor}
\usepackage{booktabs}
\usepackage{array}
\usepackage{tabularx}
\usepackage{graphicx}
\usepackage{setspace}
\usepackage{natbib}
\usepackage{hyperref}
\usepackage{xurl}
\usepackage{siunitx}
\usepackage{placeins}

\hypersetup{
  hidelinks
}

\newcommand{\E}{\mathbb{E}}
\newcommand{\Prb}{\mathbb{P}}
\newcommand{\ind}{\mathbb{I}}
\newcommand{\Z}{Z}
\newcommand{\Ztilde}{\tilde{Z}}
\newcommand{\Nhat}{\widehat N}
\newcommand{\thetahat}{\widehat\theta}
\newcommand{\Uhat}{\widehat U}

\title{When Surveys Become Conversations:\\Adaptive Matrix Validation for AI-Assisted Interviews}
\author{Tyler H. McCormick\\
Department of Statistics and Department of Sociology\\
University of Washington\\
\texttt{tylermc@uw.edu}}
\date{}

\begin{document}
\maketitle

\begin{abstract}
\begin{singlespace}

AI-assisted interviews promise to reduce respondent burden in surveys by allowing respondents to describe experiences naturally while an AI system noisily maps those accounts into structured survey variables. That mapping is a measurement process that is fallible, versioned, adaptive, and potentially behaves differently across subgroups. This paper proposes Adaptive Matrix Validation (AMV), a design in which each respondent completes an AI-assisted interview, which is then mapped into tabular data by the AI.  Respondents are also asked a small, randomized set of structured questions, which are used for statistical adjustment. The estimator first calibrates the mapped values using validation answers from other respondents, then corrects the remaining error with the validation answers observed for the target respondent. The paper develops estimators for item means, subgroup estimates, and regression coefficients when outcomes, predictors, or both are mapped from interviews. It also gives planning formulas the number of validation questions required and the sample size. A design-calibration simulation, an American Time Use Survey emulation, and a CHAMPS verbal-autopsy narrative study show when sparse validation can improve precision and when it cannot.
\end{singlespace}
\end{abstract}


\section{Introduction}
\label{sec:introduction}

With a traditional, structured survey, a person with a complex day, illness episode, political view, household arrangement, or service experience is asked to express their experience through pre-specified questions and allowed responses. The respondent has to understand what the question means, remember the relevant details, decide how those details add up, and then fit that answer into one of the survey’s available response options. \citep{tourangeau2000psychology}. The person is, in effect, fitting their experience into the tabular language that modern statistical software can understand. In other words, the burden of mapping from a complex, lived experience to tabular data falls largely upon the respondent.  AI-assisted interviews could let respondents describe their experiences conversationally while the interview system maps those accounts into structured variables for analysis, sometimes with error. This mapping has the potential to reduce survey time, improve respondent experience, and reach a broader group of individuals who may be willing to converse but not be polled (see, for example,~\citet{barari2024generative}).

This paper addresses the statistical and design challenges that arise when moving to an AI-assisted interview. Here, an AI-assisted interview is any interview in which an AI system helps ask, choose, or interpret questions during measurement.
To give a few representative examples, Xiao et al. build an AI-powered chatbot for conversational surveys that asks open-ended questions, interprets free-text answers, and probes when responses are incomplete or unclear \citep{xiao2020tell}. More recent large-language-model (LLM) systems go further: Wuttke et al. evaluate LLMs as adaptive interviewers that conduct conversational interviews on political topics \citep{wuttke2025ai}, while Barari et al. study text-based AI interviewers that dynamically probe respondents for elaboration and code open-ended answers during the survey itself \citep{barari2025ai}.

In these systems, AI selects or suggests probes, interprets natural-language responses, may apply stopping rules for follow-up questions, and maps responses into structured variables. These steps determine what was asked, what was left unasked, and how the respondent's language was mapped into structured, or tabular, data. As it performs these roles, the AI system can affect multiple error sources simultaneously. It can change comprehension by rephrasing or probing; retrieval and judgment by deciding what details to ask for; response mapping by assigning narratives to categories; processing error by extracting and coding variables; nonresponse or breakoff by changing burden and perceived privacy risk; and comparability by changing prompts, models, or stopping rules over time.  While all of these require careful attention, this paper focuses specifically on a statistical and design framework to address the measurement and processing errors introduced in an AI-assisted interview.

Specifically, this paper proposes \emph{Adaptive Matrix Validation} (AMV) as a strategy to reduce bias and quantify uncertainty for population averages, proportions, subgroup estimates, and regression coefficients. It is called adaptive matrix validation because it validates a small randomized set of items in a respondent-by-variable matrix of structured survey responses. AMV first maps the interview to a structured answer for each item. It then asks a small random set of ordinary survey questions and uses those answers to correct the mapped values. For item means and subgroup estimates with fully observed subgroup membership, the estimate averages the mapped values after a validation-based correction. For regressions, the same idea is applied to the estimating equation. This distinction matters because the same structured response variable can be an outcome in one analysis, a predictor in another, and a control or interaction in a third.
A tool that probes respondents well and maps their accounts closely to the structured variables can reduce the number of validation questions needed. When the mapped values are weak, calibration shrinks their contribution and the design relies more heavily on the validation questions themselves.

This work connects survey-statistical tools for measurement error, adaptive data collection, matrix sampling, planned missingness, computerized adaptive testing, active questionnaires, and model-assisted estimation \citep{biemer2010total,groves2010total,groves2006responsive,tourangeau2017adaptive,raghunathan1995split,graham2006planned,cochran1977sampling,sarndal1992model,zhang2020active,yoshida2023bayesian,wang2020methods} with work on conversational interviewing, machine coding of free responses, LLM coding of open-ended survey data, and generative AI in survey practice \citep{schober1997conversational,conrad2000clarifying,frisbie1968computers,mellon2024issues,halterman2025codebook,kreuter2025aapor,aapor2026responsibleai}.
AMV differs by treating the AI-assisted interview as part of the survey design.
The interview maps each item to a structured response, while randomized validation questions reveal only some of the corresponding survey answers with known item, pair, and planned-analysis probabilities. The goal is population and regression estimation tied to answers from validation questions, rather than better text coding or summarization alone. Appendix Table~\ref{tab:related} places AMV alongside adjacent literatures in more context.

The remainder of the paper proceeds as follows. Section~\ref{sec:design} presents the data, design, and methods.
Section~\ref{sec:sample-size} derives sample-size calculations for the number of structured validation questions required. Section~\ref{sec:calibration} gives a design-calibration simulation. Sections~\ref{sec:atus} and \ref{sec:champs} apply AMV to data from the American Time Use Survey (ATUS) and the CHAMPS network for global health surveillance. Section~\ref{sec:discussion} summarizes the implications and limits of the design. Appendix~\ref{app:related-literatures} expands the literature comparison, Appendix~\ref{app:baseline-estimators} gives baseline estimator forms, Appendix~\ref{app:linear-moment-regressions} gives the pairwise-moment form for linear regressions, Appendix~\ref{app:additional-simulation-atus-results} provides supporting simulation and ATUS results, Appendix~\ref{app:atus-llm-validation} reports the local-model ATUS interview-coding check, Appendix~\ref{app:champs-data} gives CHAMPS data details, and Appendix~\ref{app:disclosure} summarizes disclosure elements.

\section{Adaptive Matrix Validation}
\label{sec:design}

This section defines the data, validation-question randomization, and estimators used by AMV.

\subsection{Setting and Data}

Let \(i=1,\ldots,n\) index sampled respondents or cases, write \(\mathcal I_n\) for the sample, and let \(w_i\) be the survey weight. Let
\[
  \Z_i=(\Z_{i1},\ldots,\Z_{ip})
\]
denote the structured-response vector. This is the list of coded survey answers that a conventional questionnaire would produce, such as symptoms, activity minutes, categorical responses, time-use indicators, care-seeking variables, or derived domains. Not every item applies to every respondent, so \(E_{ij}=1\) indicates that item \(j\) applies to case \(i\). Inapplicable items are excluded from the relevant denominator rather than treated as zero. Eligibility and routing rules should be fixed before the validation answer for item \(j\) is observed. If eligibility is uncertain, the eligibility question should be asked as an anchor or included with the related validation item.

Each respondent also has fully observed background variables. Some questions are asked of everyone because they define the interview, such as the reference period, eligibility, age, sex, site, language, or diary day. Other fully observed variables are used for weights, subgroup estimates, response adjustment, or regression controls. The notation keeps these roles visible by writing core anchors as \(H_i\), analysis variables as \(W_i\), and all anchor and analysis variables available before validation assignment as \(A_i=(H_i,W_i)\). The same variable can play both roles.

The AI-assisted interview produces a natural-language record \(D_i\) and a mapped structured response \(\Ztilde_i\) saved before the relevant validation answer is observed. In a time-use interview, for example, the interview record may be mapped into minutes spent sleeping, working, and commuting. If the respondent is later asked a direct commuting question, the commuting value used in the estimator must be the one saved before that direct answer was seen. The saved value may include pre-specified coding rules, reconciliation of impossible combinations, or score-to-category mapping. If validation questions are interleaved with the interview, the item-specific value used for item \(j\) must be saved before the validation answer for item \(j\) is observed. Later text or later values that use that validation answer can be saved as post-validation outputs, but they cannot be used in the validation probability or in the mapped value corrected for that answer. AMV does not require a particular interviewer, prompt, model, or rule for follow-up questions. It requires the tool to save the mapped structured responses before validation answers are revealed and to record the information needed to reproduce the measurement: question wording, software version, questions asked, order of the conversation, stopping rule, and any recorded confidence or ``could not be determined from the interview'' labels. This paper studies how to estimate survey quantities after an interview tool has produced coded answers. Whether respondents understand, trust, or react differently to the interview must be studied separately.

Each respondent is also assigned a subset of structured validation questions. The chance of obtaining a usable validation answer may depend on what is already known, but it cannot depend on that validation answer itself. Let \(R_i=(R_{i1},\ldots,R_{ip})\), with \(R_{ij}=1\) when item \(j\)'s usable validation answer is observed and \(R_{ij}=0\) otherwise, including when the item is inapplicable, not selected, or selected but unanswered. The item probability is \(\pi_{ij}=\Prb(R_{ij}=1\mid \mathcal F_{ij},E_{ij}=1)\), where \(\mathcal F_{ij}\) is the information available when deciding whether item \(j\)'s usable validation answer will be observed. It may include anchors, paradata, earlier conversation, uncertainty scores, accumulated fieldwork and validation-coverage history, and the saved mapped value, but not the validation answer to item \(j\). If validation questions are interleaved with the interview, earlier volunteered information may enter \(\mathcal F_{ij}\) and \(\Ztilde_{ij}\); later text cannot justify \(\pi_{ij}\), though it can remain in the final record. If selected validation questions can go unanswered, \(\pi_{ij}\) should include that response process or be paired with an explicit response adjustment.

The validation answer defines the structured-response scale under the AMV protocol. Treating it as equivalent to a legacy survey item requires evidence on wording, order, context, and coding, because these features affect comprehension and response formation \citep{biemer2010total,groves2010total,tourangeau2000psychology,schober1997conversational,conrad2000clarifying}. Selected validation questions should therefore use documented wording, response options, reference periods, and coding rules. When the interview setting changes those features, the validation item should be treated as a versioned measurement protocol.

The resulting data are $\{D_i,A_i,E_i,\Ztilde_i,R_i,R_i\Z_i,\Pi_i,P_i\}$,
where \(R_i\Z_i\) denotes the observed validation answers and \(R_i\) marks which entries are observed. The object \(\Pi_i\) contains the known probabilities for individual items, item pairs, and preplanned analyses that require several answers from the same person, including response adjustments when needed. The record \(P_i\) contains the question path, timing, language, software version, coding rules, stopping rule, and uncertainty or inferability labels saved before validation answers were observed.

\subsection{Designing randomized validation tiles}

Define \(S_i\) as respondent \(i\)'s validation tile, which is the subset of applicable structured items selected for validation.
One respondent might receive symptom and care-seeking questions, while another receives timing and location questions. Across respondents, these tiles form the second-phase validation sample for the structured-response matrix. The tile is usually much smaller than the full structured questionnaire, and the choice of validation items is separate from the order in which questions are presented.  In this way, AMV uses the familiar logic of a two-phase design~\citep{bose1943samplingerror,cochran1977sampling,sarndal1992model}, combining proxies imperfectly predicted from a statistical model (in this case the AI-assisted interview tool) with strategically placed items for validation.

It also borrows from split questionnaires, matrix sampling, and planned missing data designs, where long instruments are divided across respondents to reduce burden \citep{raghunathan1995split,graham2006planned,gonzalez2008multiplematrix,thomas2006matrix,axenfeld2022split}.

For exposition, suppose first that the background variables, narrative interview, and mapped structured responses are completed before validation tiles are selected. The information available at that point is \(\mathcal F_i\). It contains \(D_i,A_i,E_i,\Ztilde_i\), uncertainty scores \(U_i=(U_{i1},\ldots,U_{ip})\), contradictions or labels saying that an answer could not be determined from the interview, accumulated fieldwork and validation-coverage history, and the path and software information available before tile selection. If validation questions are interleaved with the conversation, the same logic applies using the information available just before each validation question is selected.

The tile rule, the design that assigns a validation tile to each respondent, chooses \(S_i\) before observing the selected validation answers. In the simplest version, the study fixes \(K\), the number of non-anchor validation questions per respondent. The rule removes inapplicable items, selects \(K\) applicable items when at least \(K\) are available, and otherwise selects all applicable items. \(K\) can vary only when that variation is part of the field design, such as different burden limits by mode, language, accessibility need, or interview length. The rule should reserve some probability for broad random coverage, add support for planned subgroup reports and planned regressions, and place more validation on items, groups, languages, model versions, or fieldwork waves where the mapped values appear less reliable or previous validation data are sparse. Every item should keep at least a small chance of being validated, even when it looks easy to code. When an analysis needs two or more validated answers from the same person, the design must record the chance that those answers were asked together.

Using an AI-assisted interview requires careful consideration of both the survey design and the ultimate inferential goals. Before data collection, the analyst must ensure the validation questions are assigned in a way that gives the validation answers needed for each planned analysis.
Means and prevalences require each item to have a known positive probability of being asked and answered as a validation question for each eligible respondent, \(\pi_{ij}\). Subgroup estimates use the same item probability, and the design must provide enough validation observations within each reporting group. If subgroup membership is itself produced from the interview, the subgroup variable and outcome need same-respondent validation. Regressions require the variables in the estimating equation to be observed together. For example, if a care-seeking regression uses fever, transport, and care outside the home, those answers must sometimes be validated for the same respondent. This requirement has to be built into the field design before data collection. As the structured-response dictionary grows, the number of possible outcome, predictor, control, and interaction combinations grows quickly, so a sparse validation design cannot be expected to support every regression an analyst might later imagine. AMV therefore requires the study to name the main means, subgroup reports, item pairs, and regression blocks in advance, and the validation-tile rule must record the probabilities needed for those planned analyses. Let \(\mathcal J_q\) be the structured variables needed for planned analysis \(q\), let \(Q_{iq}=1\) when usable validation answers are observed for every variable in \(\mathcal J_q\), and define \(\pi_{iq}=\Prb(Q_{iq}=1\mid \mathcal F_i,E_{iq}=1)\), where \(E_{iq}=1\) indicates that respondent \(i\) is in the analytic population. Joint probabilities such as \(\Prb(R_{ij}=1,R_{ik}=1\mid \mathcal F_i,E_{ij}E_{ik}=1)\) are useful for item pairs, correlations, or later two-variable analyses.
For analyses added after fieldwork, the recorded design must show that the needed items could have been selected and that enough respondents actually received the relevant validation questions.

\subsection{Estimation and statistical calibration}
\label{sec:estimators}

The estimator uses two related ideas. The first is calibration. It uses the validation answers to learn how much the mapped value for item \(j\) should be trusted. If the mapped value is strongly related to the structured answer, the estimator leans on it more; if it is weak, the estimator leans on it less. This part follows an extensive literature in survey research on model-assisted estimators and double-sampling which uses information observed for everyone to improve precision, while checking it against a smaller set of higher-quality measurements \citep{cochran1977sampling,sarndal1992model}. For regression, the closest ancestor is Chen and Chen's double-sampling estimator, which uses first-phase proxy information together with second-phase validated measurements to improve regression estimates without changing the target parameter \citep{chen2000unified}. Second, after the mapped value has been calibrated, the estimator uses the validation answers to estimate what the calibrated mapped values still miss on average. It adds that missing piece back in, weighting each validation answer by the inverse of its known chance of being asked. This correction is the same basic idea as augmented inverse-probability estimation in missing-data and validation-sample problems \citep{robins1994estimation,bang2005doubly,tsiatis2006semiparametric}. Recent work on inference with predicted or imputed outcomes uses the same prediction-plus-correction logic in modern prediction settings \citep{wang2020methods,angelopoulos2023prediction,zrnic2024cross,gronsbell2024anotherlook,chen2025unified,salerno2025modern,salerno2025moment}.

This subsection begins with individual item means/prevalences, then moves to subgroups and, finally, regressions.
For item \(j\), calibration starts with the mapped answer \(\Ztilde_{ij}\) and pulls it toward the validation-question mean estimated from other folds. The amount of pull is governed by \(\lambda\). When the mapped answers closely match validation answers in the training folds, \(\lambda\) is closer to one. When they do not, \(\lambda\) is closer to zero. This paper uses a simple calibration because sparse validation tiles can make more detailed item-specific models unstable. A respondent's validation answer can correct the final estimate, but it should not also decide how much to use that respondent's mapped answer. To keep those roles separate, partition the respondents into \(K\) groups, called folds, and let \(k(i)\) denote the fold containing respondent \(i\). Then, for fold \(k\), define the item-specific mean:
\[
  \widehat\mu_{j,-k}
  =
  \frac{
    \sum_{\ell:k(\ell)\ne k}
    w_\ell E_{\ell j}R_{\ell j}\Z_{\ell j}/\pi_{\ell j}
  }{
    \sum_{\ell:k(\ell)\ne k}
    w_\ell E_{\ell j}R_{\ell j}/\pi_{\ell j}
  }.
\]
For respondent \(i\), define the calibrated response to item \(j\) by
\[
  M_{ij}^{(-k(i))}
  =
  \widehat\mu_{j,-k(i)}
  +
  \widehat\lambda_{j,-k(i)}
  \{\Ztilde_{ij}-\widehat\mu_{j,-k(i)}\},
  \qquad
  \widehat\lambda_{j,-k(i)}\in[0,1]
\]

In this simple version, \(\widehat\lambda_{j,-k(i)}\) controls how much the mapped answer can contribute to precision. A value of 1 leaves \(\Ztilde_{ij}\) unchanged.
Choosing \(\lambda\) matters because calibration alone can reduce bias, while interval-width protection comes from using the mapped answer only when the training-fold validation answers show that it helps precision. More detailed calibration rules can use uncertainty scores, inferability labels, question-path features, or subgroup features, but they must be learned from other folds rather than from the held-out respondent's validation information.

The value of \(\lambda\) is chosen in the training folds. For each candidate value between 0 and 1, compute the calibrated mapped answer implied by that value and compare it with the observed validation answers in the training folds:
\[
  \widehat Q_{j,-k}(\lambda)=
  \sum_{\ell:k(\ell)\ne k}
  \frac{R_{\ell j}}{\pi_{\ell j}}w_\ell^2E_{\ell j}
  \left(\frac{1}{\pi_{\ell j}}-1\right)
  \left[
    \Z_{\ell j}
    -
    \{\widehat\mu_{j,-k}+\lambda(\Ztilde_{\ell j}-\widehat\mu_{j,-k})\}
  \right]^2.
\]
The chosen value \(\widehat\lambda_{j,-k}\in\arg\min_{\lambda\in[0,1]}\widehat Q_{j,-k}(\lambda)\) is the value that gives the smallest estimated validation-assignment variance for this item. To see where the weights come from, fix a calibrated value \(M_{ij}\) and write the remaining error as \(e_i=\Z_{ij}-M_{ij}\). The validation-weighted correction contains \(R_{ij}e_i/\pi_{ij}\). Under the diagonal validation-assignment calculation, with sampled records held fixed,

\[
  \operatorname{Var}_R\left\{
    w_iE_{ij}\frac{R_{ij}}{\pi_{ij}}e_i
    \mid \mathcal F_{ij},\Z_{ij},E_{ij}=1,M_{ij}
  \right\}
  =
  w_i^2E_{ij}
  \left(\frac{1}{\pi_{ij}}-1\right)e_i^2.
\]
Survey weights enter squared because this is a variance calculation. The factor \(R_{\ell j}/\pi_{\ell j}\) in \(\widehat Q_{j,-k}\) appears because \(e_\ell^2\) is observed only for respondents who answered item \(j\) as a validation question, and those observed residuals must represent the full eligible training-fold population. This criterion is the training-fold version of the conditional validation-assignment variance for a fixed calibrated value. Because \(\widehat\mu_{j,-k}\) is itself estimated in the training folds, an inner cross-fit version can be used when an exactly fold-external tuning criterion is desired. If the training folds contain too few usable validation answers for item \(j\), the analysis must use a pre-specified fallback, such as a previous wave, a pooled estimate across related items, or \(\widehat\lambda_{j,-k}=0\), or report that the item lacks enough validation support for calibrated AMV. When selected items are linked within fixed-size tiles or blocks, the same idea should use the relevant covariance terms, pair or block probabilities, replicate weights, or a nested resampling analogue.

After calibration, the estimator uses the validation answers to correct for remaining error. The calibrated value $M_{ij}^{(-k(i))}$ is a working approximation
  available for every respondent. Among respondents who answered item $j$ as a validation question, the gap between their validation answer and their calibrated
  value, $Z_{ij} - M_{ij}^{(-k(i))}$, is observed. These gaps reveal what the calibration still misses. Because only some respondents receive each validation
  question, the estimator reweights each observed gap by the inverse of its known selection probability, $1/\pi_{ij}$, so that the correction applies to the full
  eligible population rather than just the validated subset:

\begin{equation}
  \thetahat^{cal}_j =
  \frac{1}{\Nhat_j}
  \sum_{i\in\mathcal I_n}w_i
  E_{ij}
  \left[
    M_{ij}^{(-k(i))}
    +
    \frac{R_{ij}}{\pi_{ij}}
    \{\Z_{ij}-M_{ij}^{(-k(i))}\}
  \right].
  \label{eq:item-cal-amv}
\end{equation}
For any calibrated mapped value fixed before respondent \(i\)'s own validation assignment, response status, and validation answer for item \(j\) can affect it, the known-probability validation assignment gives
\begin{equation}
  \E_R\left[
    M_{ij}^{(-k(i))}
    +
    \frac{R_{ij}}{\pi_{ij}}\{\Z_{ij}-M_{ij}^{(-k(i))}\}
    \mid \mathcal F_{ij},\Z_{ij},E_{ij}=1,M_{ij}^{(-k(i))}
  \right]
  =
  \Z_{ij}.
  \label{eq:item-cal-id}
\end{equation}
The expectation is over the validation-tile assignment, holding the sampled records and their structured responses fixed. Under those conditions, calibration can reduce noise, while the validation-weighted correction keeps the estimate tied to the structured-response target.

Let \(G_i\) denote a subgroup label observed before validation assignment, such as sex, age group, site, or language. For subgroup \(g\), the corresponding estimator is:
\begin{equation}
  \thetahat^{cal}_{jg} =
  \frac{
    \sum_{i\in\mathcal I_n} w_i \ind(G_i=g)E_{ij}
    \left[
      M_{ij}^{(-k(i))}
      +
      R_{ij}\{\Z_{ij}-M_{ij}^{(-k(i))}\}/\pi_{ij}
    \right]
  }{
    \sum_{i\in\mathcal I_n} w_i \ind(G_i=g)E_{ij}
  }.
  \label{eq:subgroup-cal-amv}
\end{equation}
If a priority subgroup has few validation-question answers for item \(j\), the estimate will be noisy and any remaining error will be hard to see. The validation rule must assign enough validation questions within subgroups that will be reported. If subgroup membership is itself a mapped structured response item, then subgroup estimation requires the subgroup item and \(\Z_j\) to be observed through validation questions for the same respondent. That case needs a version that validates both the subgroup label and the outcome for the same respondent.

For regressions, AMV uses the same order as the item estimator, but applies it to an estimating equation. The design requirement is stronger than for a single item. A regression score can combine outcomes, predictors, controls, interactions, and transformations, so the structured answers needed to evaluate that score must sometimes be validated together for the same respondent. Separate validation of the outcome for one respondent and a predictor for another is not enough for this score form.

For a planned regression \(q\), let \(\mathcal J_q\) be the structured responses needed to evaluate its score, and let \(\beta\) be the coefficient vector for that regression. Define the unweighted complete-data score contribution
\[
  \psi_i(\beta)=\psi_q(\Z_i,A_i;\beta),
\]
so that the planned weighted estimating equation would use the sample sum of \(w_iE_{iq}\psi_i(\beta)\). Let
\[
  \widetilde\psi_i(\beta)=\psi_q(\Ztilde_i,A_i;\beta)
\]
be the same score evaluated from the mapped record. This setup covers estimating equations whose complete-data score can be evaluated from a same-respondent validation block, including interactions, transformations, outcomes, predictors, and controls when the needed variables are observed together \citep{robins1994estimation}. Here \(E_{iq}=1\) means respondent \(i\) is in the analytic population for regression \(q\) and the variables in \(\mathcal J_q\) apply. Let \(Q_{iq}=1\) when usable validation answers are observed for every structured response in \(\mathcal J_q\), and set \(Q_{iq}=0\) otherwise, including when \(E_{iq}=0\). Let \(\mathcal F_{iq}\) be the information available before the validation block for \(\mathcal J_q\) is assigned; when validation tiles are selected after the interview, this is the corresponding block version of \(\mathcal F_i\). Among applicable analytic records, \(\pi_{iq}=\Prb(Q_{iq}=1\mid \mathcal F_{iq},E_{iq}=1)\) is the known probability that the usable same-respondent validation block is observed, with \(0<\pi_{iq}\le 1\). For inapplicable records, the factor \(E_{iq}\) makes the contribution zero.

Calibration first chooses a mapped score for each respondent. Let \(C_{iq}^{(-k(i))}(\beta)\) be that calibrated mapped score. It can use the mapped record, anchors, saved uncertainty scores, inferability labels, and other information logged before the needed validation answers are observed. It cannot use respondent \(i\)'s own validation assignment, response status, or validation answers for regression \(q\). A simple first version uses one multiplier for the mapped regression score:
\[
  C_{iq}^{(-k(i))}(\beta)=\widehat\lambda_{q,-k(i)}\widetilde\psi_i(\beta),
  \qquad \widehat\lambda_{q,-k(i)}\in[0,1].
\]
This is close in spirit to Chen and Chen's double-sampling regression estimator~\citep{chen2000unified}. In their setting, rough or proxy measurements are observed for the full primary sample and exact measurements are observed for a simple random validation subsample. They start from the validation-sample regression estimate and improve it by adding a matrix-weighted difference between a working-model estimate from the validation subsample and the corresponding estimate from the full primary sample. That matrix is built from the large-sample covariance structure and gives the most efficient constant-matrix adjustment in their class \citep{chen2000unified}. AMV uses the same broad idea: proxy information improves precision, while validation answers keep the target on the structured-response scale. The implementation here uses a simpler scalar score calibration rather than a matrix adjustment. That choice is deliberate. Sparse validation tiles often leave only a limited number of respondents with the full validation block for a planned regression, especially when the regression uses several mapped variables. A single multiplier is easier to pre-specify, easier to estimate with limited block support, and easier to report.

To choose \(\lambda\), start with a preliminary coefficient without the held-out fold. It can come from validation blocks in the training folds, a previous wave, a pilot, or a pre-specified training-fold AMV estimate. If validation assignment or sampling is clustered, the folds should be formed at that level. At a preliminary value \(\widehat\beta^{(0)}_{q,-k}\), the training-fold rule chooses \(\lambda\) by comparing the validated score with the \(\lambda\)-scaled mapped score:
\[
  \widehat Q_{q,-k}(\lambda)=
  \sum_{i:k(i)\ne k}
  \frac{Q_{iq}}{\pi_{iq}}w_i^2E_{iq}
  \left(\frac{1}{\pi_{iq}}-1\right)
  \left\|
    \psi_i(\widehat\beta^{(0)}_{q,-k})
    -
    \lambda\widetilde\psi_i(\widehat\beta^{(0)}_{q,-k})
  \right\|^2 .
\]
The fitted multiplier is \(\widehat\lambda_{q,-k}\in\arg\min_{\lambda\in[0,1]}\widehat Q_{q,-k}(\lambda)\). Values near one give more weight to the mapped score. Values near zero make the final equation rely mainly on same-respondent validation blocks. The weights in \(\widehat Q_{q,-k}\) have the same source as the item weights. For \(r_i(\lambda)=\psi_i(\widehat\beta^{(0)}_{q,-k})-\lambda\widetilde\psi_i(\widehat\beta^{(0)}_{q,-k})\), the validation-assignment part of the score correction has diagonal variance contribution proportional to
\[
  w_i^2E_{iq}
  \left(\frac{1}{\pi_{iq}}-1\right)
  \|r_i(\lambda)\|^2 .
\]
The factor \(Q_{iq}/\pi_{iq}\) estimates this quantity from training-fold records where the full validation block was observed for the same respondent. The norm should be chosen before analysis. Ordinary squared differences tune the score on its own scale; a pre-specified matrix or norm is needed when the goal is to put more weight on selected coefficients or contrasts. Recent reference-sample methods use related tuning ideas for fitted mapped values \citep{angelopoulos2023prediction,zrnic2024cross}.

After calibration, the validation blocks estimate what the calibrated mapped score still misses. For any calibrated mapped score fixed before respondent \(i\)'s own validation block can affect it, the corrected score starts with \(C_{iq}^{(-k(i))}(\beta)\) and adds back the validation-weighted difference between the validated score and the calibrated mapped score:
\begin{equation}
  \Uhat_q(\beta;C)=
  \sum_{i\in\mathcal I_n} w_iE_{iq}
  \left[
    C_{iq}^{(-k(i))}(\beta)
    +
    \frac{Q_{iq}}{\pi_{iq}}
    \{\psi_i(\beta)-C_{iq}^{(-k(i))}(\beta)\}
  \right].
  \label{eq:score-cal-amv}
\end{equation}
The calibrated AMV regression estimate solves \(\Uhat_q(\beta;C)=0\). For a calibrated mapped score fixed without respondent \(i\)'s own score-validation block, the known validation probability gives
\begin{equation}
  \E_R\left[
    C_{iq}^{(-k(i))}(\beta)
    +
    \frac{Q_{iq}}{\pi_{iq}}
    \{\psi_i(\beta)-C_{iq}^{(-k(i))}(\beta)\}
    \mid \mathcal F_{iq},\Z_i,A_i,E_{iq}=1,C_{iq}^{(-k(i))}(\beta)
  \right]
  =
  \psi_i(\beta).
  \label{eq:score-cal-id}
\end{equation}
The weighted estimating equation therefore targets the same planned structured-response score that would be available if the needed structured responses were observed for every applicable sampled record. With the scalar calibrated mapped score above, the final equation is
\[
  \sum_{i\in\mathcal I_n} w_iE_{iq}
  \left[
    \widehat\lambda_{q,-k(i)}\widetilde\psi_i(\beta)
    +
    \frac{Q_{iq}}{\pi_{iq}}
    \{\psi_i(\beta)-\widehat\lambda_{q,-k(i)}\widetilde\psi_i(\beta)\}
  \right]
  =0.
\]
If a training fold lacks enough same-respondent validation blocks, the analysis must use a pre-specified fallback, such as a previous wave, a pilot, or a pooled training estimate, or report that the planned regression lacks support for that fold.
Appendix~\ref{app:linear-moment-regressions} gives the corresponding pairwise-moment form for linear regressions.

The same corrected quantities are used for uncertainty estimates. For item means, define the calibrated pseudo-outcome
\[
  \phi_{ij}=E_{ij}
  \left[
    M_{ij}^{(-k(i))}
    +
    R_{ij}\{\Z_{ij}-M_{ij}^{(-k(i))}\}/\pi_{ij}
  \right].
\]
If the sample is treated as independent after weighting, a first-order variance estimator is
\[
  \widehat V(\thetahat^{cal}_j)=
  \frac{1}{\Nhat_j^2}
  \sum_{i\in\mathcal I_n} w_i^2(\phi_{ij}-E_{ij}\thetahat^{cal}_j)^2,
\]
with the usual stratum and cluster modifications for complex samples. Standard errors should, of course, reflect both the original survey design and the random validation questions. When the survey has weights, strata, clusters, fixed-size validation tiles, validation blocks, or adaptive assignment, the variance calculation should use those features \citep{cochran1977sampling,sarndal1992model}. Replicate weights are preferred when they are available. If the calibration parameter is estimated from the validation data and has a meaningful effect on the reported interval, it should be re-estimated inside each replicate or bootstrap sample when feasible.

\section{Sample Size and Feasibility}
\label{sec:sample-size}

This section outlines a framework for understanding how changes to AMV design impact the (expected) variance of the estimators, thereby giving a sense of when using a particular AI-assisted interview tool would increase efficiency over a traditional survey. AMV depends on three quantities. The first is how well the interview record \(D_i\) predicts the structured response. Better mapping leaves less error for validation questions to correct. The second is the number of validation questions per respondent, denoted by \(B\). The third is the effective sample size. A design can sometimes trade one for another. A stronger mapping can support fewer validation questions. More validation questions can compensate for a weaker mapping. A larger effective sample can make either design more precise. These tradeoffs are partial, not automatic, because rare items, routing, uneven weights, and planned regressions all need their own validation support.

For item \(j\), let \(n_{\mathrm{eff},j}\) be the effective number of applicable cases, let \(\sigma_j^2=\mathrm{Var}(\Z_{ij}\mid E_{ij}=1)\), and let \(q_j=\Prb(R_{ij}=1\mid E_{ij}=1)\) be the planning probability that an applicable respondent receives item \(j\) as a validation question. If selected validation questions can go unanswered, \(q_j\) should include that response process or be paired with a response adjustment.

The mapping-quality input is how much item-level variation remains unexplained after calibration. For a mapped value \(M_{ij}\), either raw or calibrated, define
\begin{equation}
  \rho_j(M) =
  \frac{\E\{(\Z_{ij}-M_{ij})^2\mid E_{ij}=1\}}{\mathrm{Var}(\Z_{ij}\mid E_{ij}=1)}.
  \label{eq:rho}
\end{equation}
This ratio compares the error left after using the mapped value with the natural variation in the structured item. When \(M_{ij}=\Ztilde_{ij}\), it describes the raw mapped value. When \(M_{ij}\) is the fold-external calibrated mapped value, it describes the calibrated AMV input. It is often easier to read this as \(1-\rho_j(M)\): the share of item-level variation explained by the mapped value, relative to using only the item mean. A value of \(\rho_j(M)=0.10\) means the mapped value explains about 90 percent of the item-level variation and leaves 10 percent for validation questions to correct. A value near one means the mapping gives little precision gain. For a binary item, the denominator is \(p_j(1-p_j)\).

For planning, use a conservative unexplained-variation ratio \(\rho_j^{plan}\). When validation support is thin, set \(\rho_j^{plan}=1\). Otherwise, use a capped upper bound of the estimated calibrated unexplained-variation ratio, such as
\[
  \rho_j^{plan}=\min\{1,\widehat\rho_j^\star+c\,\widehat{SE}(\widehat\rho_j^\star)\},
\]
where \(\widehat\rho_j^\star\) is the estimated unexplained-variation ratio after calibration and \(c\) is chosen before fielding. The cap at one reflects the separate benchmark that uses only validation questions: if the calibrated mapping appears worse than that benchmark for planning purposes, do not plan on a precision gain from it.

If each applicable respondent has the same validation probability \(q_j\) for item \(j\), a useful planning approximation for variance is
\begin{equation}
  \mathrm{Var}(\thetahat^{cal}_j)
  \approx
  \frac{\sigma_j^2}{n_{\mathrm{eff},j}}
  \left[
    1+\left(\frac{1}{q_j}-1\right)\rho_j^{plan}
  \right].
  \label{eq:variance-approx}
\end{equation}
The first factor is ordinary sampling uncertainty with complete structured responses. The bracket is the extra uncertainty from giving item \(j\) as a validation question to only some applicable respondents. When \(q_j=1\), the bracket is one. When \(q_j<1\), sparse validation is affordable only to the extent that the calibrated mapping leaves little unexplained variation.

The same calculation can be read as a sample-size requirement. Suppose the target margin of error is \(h_j\). For a proportion, a 95 percent interval with half-width 5 percentage points uses \(h_j=0.05\) and \(z=1.96\). The required effective sample size is approximately
\begin{equation}
  n_{\mathrm{eff},j}
  \ge
  \frac{z^2\sigma_j^2}{h_j^2}
  \left[
    1+\left(\frac{1}{q_j}-1\right)\rho_j^{plan}
  \right].
  \label{eq:n-required}
\end{equation}
This is the main planning formula. It says that the complete-structured-response sample size is multiplied by an extra factor. That factor is small when the mapping is strong or the item is often included in the validation tile. It is large when the mapping is weak and the item is rarely validated.

If every item applies to everyone and validation questions are spread roughly evenly across \(p\) items, then \(q_j\approx B/p\). In that simple case,
\begin{equation}
  n_{\mathrm{eff},j}
  \ge
  \frac{z^2\sigma_j^2}{h_j^2}
  \left[
    1+\left(\frac{p}{B}-1\right)\rho_j^{plan}
  \right].
  \label{eq:n-required-b}
\end{equation}
This version shows the tradeoff. For a fixed precision target, the study can increase \(B\), improve the mapping, increase the sample size, or relax the target. None of these choices is free. More validation questions increase respondent burden. Improving the mapping requires a better interview protocol or coding rule. Increasing \(n_{\mathrm{eff},j}\) requires more usable sample. \(p\) counts structured fields after routed subparts are expanded. It is larger than the number of printed questionnaire prompts. For scale, the ACS household form has 27 numbered housing questions and a Person 1 section that runs through 44 numbered person questions. Census estimates the average household completion time at 40 minutes~\citep{CensusACS2025Questionnaire}. After expanding subparts into structured fields, a one-person ACS-like instrument is on the order of 120 structured items. The example below therefore uses \(p=120\).

\begin{figure}[t]
\centering
\includegraphics[width=0.82\textwidth]{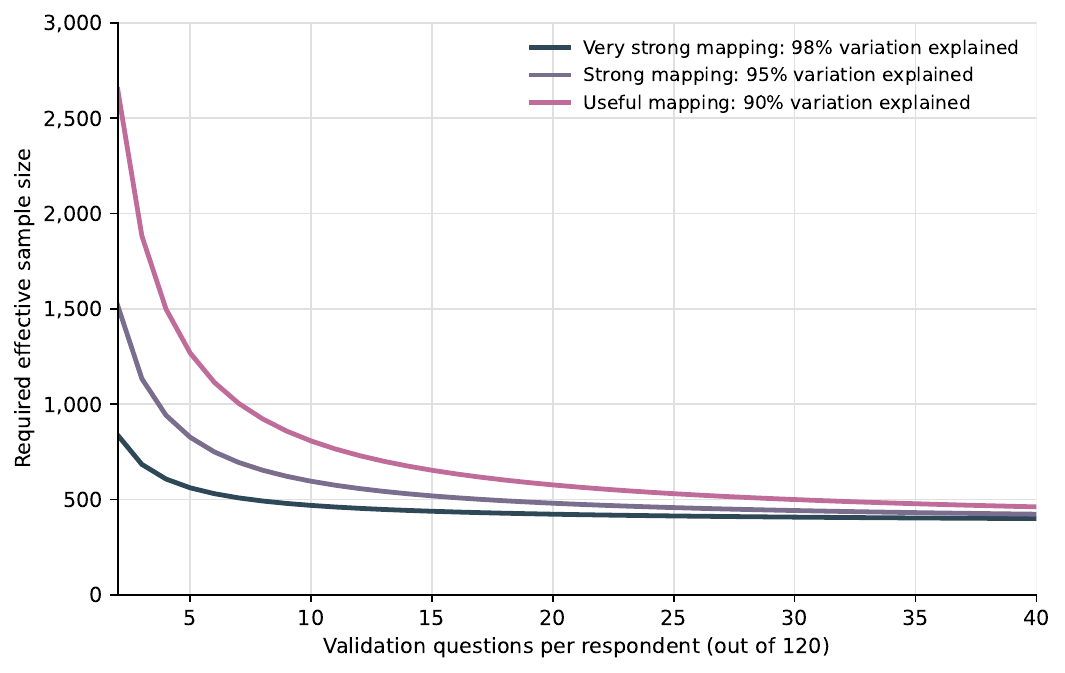}
\caption{Planning illustration for the tradeoff among mapping quality, validation questions, and effective sample size. The figure uses hypothetical values, not empirical results: \(p=120\) binary items, \(\sigma^2=0.25\), target margin of error \(h=0.05\), and \(z=1.96\). The value \(p=120\) is roughly the size of a one-person ACS-like structured-item universe after expanding question subparts into fields the study wants available for analysis. The three lines show high-quality mappings after calibration, where the mapped value explains 98, 95, or 90 percent of the item-level variation relative to using only the item mean. These are realistic planning values for items that the interview probes directly or that are clearly described in the respondent's account. Required effective sample size falls when each respondent receives more validation questions, and it falls faster when the mapped value explains more of the item-level variation.}
\label{fig:sample-size-tradeoff}
%\figalt{Line chart showing required effective sample size against validation questions per respondent for mapped values that explain 98, 95, and 90 percent of item-level variation. The 90 percent line is highest, the 95 percent line is in the middle, and the 98 percent line is lowest; all three decline as the number of validation questions increases.}
\end{figure}

Figure~\ref{fig:sample-size-tradeoff} is an illustration. It focuses on settings where AMV could reduce burden because the mapped values are strong enough for sparse validation to improve precision without an unrealistically large sample. If early validation data suggest that the unexplained share \(\rho_j^{plan}\) is much larger than these values, the same formula gives a practical warning. The study should ask that item more often, improve the interview so it probes the content more directly, increase the sample size, or treat the item as one that needs ordinary structured questioning.

Sometimes the study instead starts with a fixed sample size and asks how often item \(j\) must be validated. Define
\[
  m_j=\frac{n_{\mathrm{eff},j}h_j^2}{z^2\sigma_j^2}.
\]
The number \(m_j\) says how much room the target leaves for not asking item \(j\) of everyone. If \(m_j=1\), the target is as tight as the interval would be if every applicable respondent answered that structured item. If \(m_j>1\), the target allows some extra uncertainty. Solving \eqref{eq:variance-approx} gives
\begin{equation}
  q_j \ge
  \frac{\rho_j^{plan}}{m_j-1+\rho_j^{plan}}.
  \label{eq:q-required}
\end{equation}
A value of \(q_j=0.06\) means asking the item as a validation question of about 6 percent of applicable respondents. If \(m_j<1\), the requested margin of error is tighter than this sample can reach, even with a perfect mapping. Ordinary survey sampling error would still exceed the target if every applicable respondent answered item \(j\). The design should still keep a pre-specified minimum validation probability for every item, because mapping accuracy can differ across groups, languages, field periods, or versions of the interview protocol.

The item-level probabilities add up to the average number of non-anchor validation questions per respondent. If every item applies to every respondent,
\[
  B_{\mathrm{required}}=\sum_{j=1}^p q_j.
\]
With routing, each item is weighted by the share of respondents to whom it applies:
\[
  B_{\mathrm{required}}
  \approx
  \sum_{j=1}^p \bar E_j q_j,
\]
where \(\bar E_j\) is the population share for whom item \(j\) applies. This count covers only item-by-item precision. Core anchors, same-respondent validation for planned regressions, minimum validation for important subgroups, and allowances for unanswered validation questions must be added when they are part of the design.

Realized validation support should be reported, not only planned. For item \(j\), define validation weights \(v_{ij}=w_iE_{ij}R_{ij}/\pi_{ij}\). A useful effective validation sample size is
\[
  n_{\mathrm{eff},j}^{val}=
  \frac{\left(\sum_i v_{ij}\right)^2}{\sum_i v_{ij}^2}.
\]
For binary items, also report effective positives and negatives. Claims should be weaker when support is low, outcomes are rare, propensities are very small, or a few weights dominate, even if the calibrated estimate looks precise. Subgroup claims require the same support inside the subgroup, not only in the full sample.

Regressions need overlap, not only coverage. To estimate a mean for item \(j\), the design needs enough validation answers for item \(j\). To estimate a regression, it needs the outcome and predictors validated for the same respondents. If each respondent receives \(B\) validation items chosen at random from \(p\) items, the chance that any specified pair is validated together is
\[
  q_2=\frac{B(B-1)}{p(p-1)}.
\]
The expected effective number of respondents with both answers is \(n_{\mathrm{eff}}q_2\). For a larger item universe, say \(p=420\) and \(n_{\mathrm{eff}}=5{,}000\), random selection needs about \(B=43\) validation questions per respondent to get about 50 effective observations for a given pair. This is why planned regressions should be built into the question-selection rule instead of left to random overlap.

For a planned regression indexed by \(q\), use the effective number of applicable records and the probability that all variables needed for the score are validated together. With \(v_{iq}=w_iE_{iq}Q_{iq}/\pi_{iq}\), the realized same-respondent support is
\[
  n_{\mathrm{eff},q}^{val}=
  \frac{\left(\sum_i v_{iq}\right)^2}{\sum_i v_{iq}^2}.
\]
Before fielding, use \(n_{\mathrm{eff},q}\bar\pi_q\), where \(\bar\pi_q\) is the design-average probability that the validation tile contains all variables needed for the score. The residual variation that matters is the calibrated score residual \(\psi_i(\beta)-C_{iq}^{(-k(i))}(\beta)\), not only item-by-item mapping error. The design implication is simple. Broad item coverage supports means and proportions. Planned subgroup reports need validation inside the subgroup. Planned regressions need same-respondent validation for the variables in the score. Sparse overlap across unrelated topics can help later analysis, but it does not support regressions whose score variables were not validated for the same respondents.

\section{Simulation}
\label{sec:calibration}

Before the empirical examples, a small simulation checks the basic behavior of the estimators. It creates \(n=5{,}000\) respondents and repeats 800 times. In the item-mean setting, a binary structured response depends on a covariate and subgroup, and the mapped value is informative but underestimates the subgroup effect. Each respondent receives the target validation question with probability \(q\). In the regression setting, the outcome and one predictor are noisily mapped from the interview record, and with probability \(q\) the validation tile contains both variables for the same respondent. In both settings, the target is the finite-population quantity that would be computed from the complete structured responses. Appendix~\ref{app:baseline-estimators} gives the validation-question-only baseline estimators used for comparison.

\begin{figure}[t]
\centering
\includegraphics[width=0.95\textwidth]{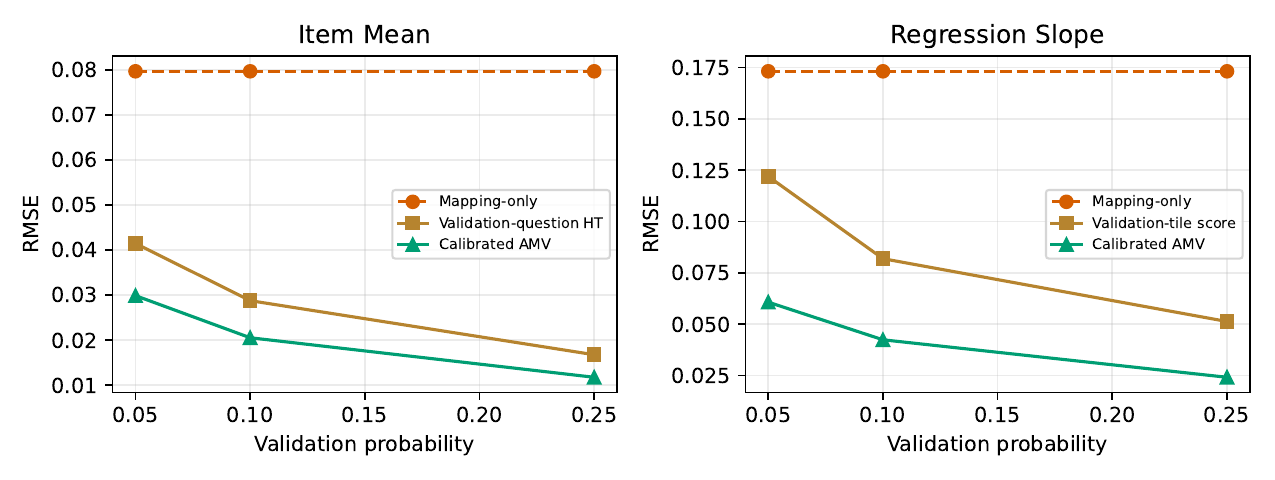}
\caption{Validation probability and root mean squared error in the design-calibration simulation. The panels show item-mean and regression-score settings over validation probabilities \(q=0.05,0.10,0.25\) in 800 repetitions with \(n=5{,}000\). Mapping-only error stays flat because it does not use validation answers, while estimates using only validation questions and AMV improve as validation probability increases; AMV uses the informative mapped value for precision.}
\label{fig:calibration}
%\figalt{Two-panel line chart comparing RMSE across validation probabilities. Mapping-only estimation remains high and flat, while estimates using only validation questions and AMV improve as validation probability increases; AMV is lowest when the mapped value is informative.}
\end{figure}

Figure~\ref{fig:calibration} shows RMSE for validation probabilities of 0.05, 0.10, and 0.25. Mapping-only estimation remains biased when the mapped values have systematic error, because it has no validation answers to correct that error. Validation questions only and AMV both improve as \(q\) increases, but AMV has lower RMSE when the mapped value is informative and stays close to the finite-population target. For \(q=0.10\), Appendix Table~\ref{tab:calibration} reports bias, RMSE, and coverage. The next sections evaluate the methods in an ATUS emulation and a CHAMPS narrative study.

\section{ATUS Matrix-Validation Study}
\label{sec:atus}

The American Time Use Survey (ATUS) is a nationally representative U.S. time-diary survey that asks respondents age 15 and older to report activities over a 24-hour day, with activities coded into a structured lexicon and released with survey weights for population estimates \citep{BLS2025ATUSMicrodata,HamermeshFrazisStewart2005ATUS}.
This experiment uses public ATUS respondent and activity files for 2018, 2019, and 2021--2024. The analytic sample includes 52,468 respondent-days and 970,712 activity episodes.

For each respondent-day, 32 variables are derived from the complete diary and form \(Z_i\). This reference retains the usual recall, reporting, coding, and processing errors of a time-use survey.
The experiment creates a synthetic \(\Ztilde_i\) for every respondent and all 32 variables. The error introduced by filling in the structured items from the AI-assisted interview is random and varies across three settings. Strong, moderate, and weak scenarios differ in error and systematic bias, with larger errors for domains that are plausibly hard to infer from a short narrative, like commute and other short travel, direct and secondary childcare, work-at-home distinctions, and the boundary between screen time and other leisure.  While this strategy makes the experiment semi-synthetic, it also facilitates isolating the types of errors AMV is designed to correct. Running full interviews would make it harder to tell whether AMV corrects the estimates, because the results would also depend on question wording, probing, coding, and software behavior. The size of these errors was calibrated with a small LLM experiment. One LLM played the interviewer, using ATUS diary facts to ask fixed questions. A second LLM played the respondent and answered in natural language. A third LLM coded those answers back into ATUS variables. Appendix~\ref{app:atus-llm-validation} gives the details.

Validation assignment rules are defined over 250 possible validation items, including the 32 reported ATUS variables plus 218 additional diary checks. The 218 additional checks count toward respondent burden and affect the validation assignment, even though they are not used in the reported means or regressions.
The validation tiles contain \(B=12,18,\) or \(25\) structured items per respondent-day over 30 simulation repetitions. The results presented here for regression use \(B=18\), or 7.2 percent of the validation universe. In the moderate-error setting, each priority item is directly validated for about 7.4 percent of respondent-days, and the two planned regression blocks are validated for about 6.0 and 6.1 percent. This gives roughly 2,100 effective item-validation observations and about 1,700 effective same-respondent regression-block observations, far below the full sample of 52,468 respondent-days.
Most validation questions came from extra diary checks, but the design also reserved some questions for broad random coverage and some for the variables needed in the planned regressions. This made sure that enough respondents answered the outcome and predictor validation questions together.

\begin{figure}[h]
\centering
\includegraphics[width=.85\textwidth]{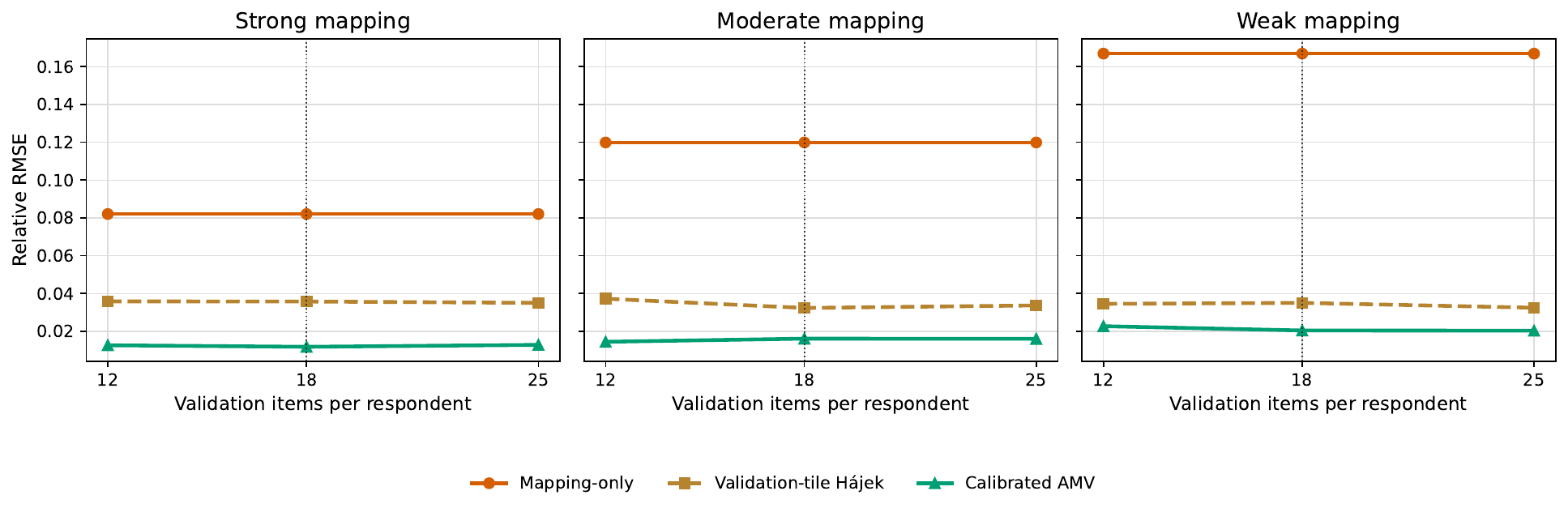}
\caption{ATUS error by number of validation items. The y-axis is mean RMSE divided by the absolute complete-diary reference value over seven priority variables: sleep, paid work, commute, screen time, exercise, secondary childcare, and direct-childcare participation. Lower values are better. Each panel shows mapping-only estimation, validation-tile H\'ajek, and calibrated AMV for one simulated error setting, with common y-axis limits across panels and a vertical dotted line marking the main \(B=18\) setting. Calibrated AMV is the fold-external calibrated estimator with a tuned mapped-control term. Validation is assigned over 250 possible items, so \(B=18\) corresponds to 7.2 percent of the validation universe. Results are from the semi-synthetic experiment rather than a deployed AI interviewer.}
\label{fig:atus-frontier}
%\figalt{Three-panel line chart for strong, moderate, and weak simulated error settings. Mapping-only estimates have larger error when the simulated mapped value is biased. Validation questions only and calibrated AMV generally improve as more validation items are revealed, with smaller gains when the mapped value is weak.}
\end{figure}

Figure~\ref{fig:atus-frontier} reports RMSE for \(B=12,18,\) and \(25\) validation items under the strong, moderate, and weak error settings. The figure shows two patterns. Mapping-only estimates (what would happen if the researcher used the output of the AI-assisted interview naively) carry the bias built into the simulated mapped values. Validation-tile H\'ajek and calibrated AMV improve as \(B\) increases, because a larger tile raises the validation probability for target items and gives more observations for estimating mapped-value errors. The calibrated AMV item estimator uses a calibrated mapped value derived from folds that don't include the respondent and then tunes how much to use the mapped term. The average item \(\lambda\) values for the seven priority variables are close to one in the \(B=18\) setting, corresponding to a highly effective AI-assisted interview. In this 32-variable recode, with the 250-item validation design, the simulated error mechanism, and the logged validation propensities, the calibrated AMV item biases are near zero for the priority variables. The validation-tile H\'ajek benchmark is also centered near the reference value, but it is noisier for these item means because it discards the mapped values observed for everyone. Across the seven items, three error settings, and three validation burdens, calibrated AMV has lower RMSE than validation-tile H\'ajek in 62 of 63 comparisons and a smaller linearized standard-error approximation in all 63 comparisons. Appendix Table~\ref{tab:atus-items} lists supporting item-level numerical values and additional simulation details.

For regression, the first example is a sleep-minutes regression motivated by time-use work on sleep and waking activities \citep{basner2007sleep}. The structured outcome is sleep minutes (nightly). The mapped structured predictors are paid-work minutes, commute minutes, and screen-time minutes. Age, sex, education, employment, weekend diary day, children in the household, and survey year are observed background variables for all respondents. The second regression is a linear-probability model for any direct childcare, motivated by the parental time-use literature on employment, gender, education, and household structure \citep{guryan2008parental,raley2012fathers,pepin2018marital}. Its structured outcome is the direct-childcare participation indicator. The mapped structured predictor is paid-work minutes. The same background covariates are included. The childcare coefficient is children in the household, an always-observed, anchor predictor. This makes the childcare panel mainly about correction of the mapped childcare outcome. The paid-work coefficient in this same regression is not shown because its mapping-only bias is close to zero through offsetting errors in mapped childcare and mapped paid work, rather than because the mapped regression is generally accurate.

\begin{figure}[p]
\centering
\includegraphics[width=0.75\textwidth]{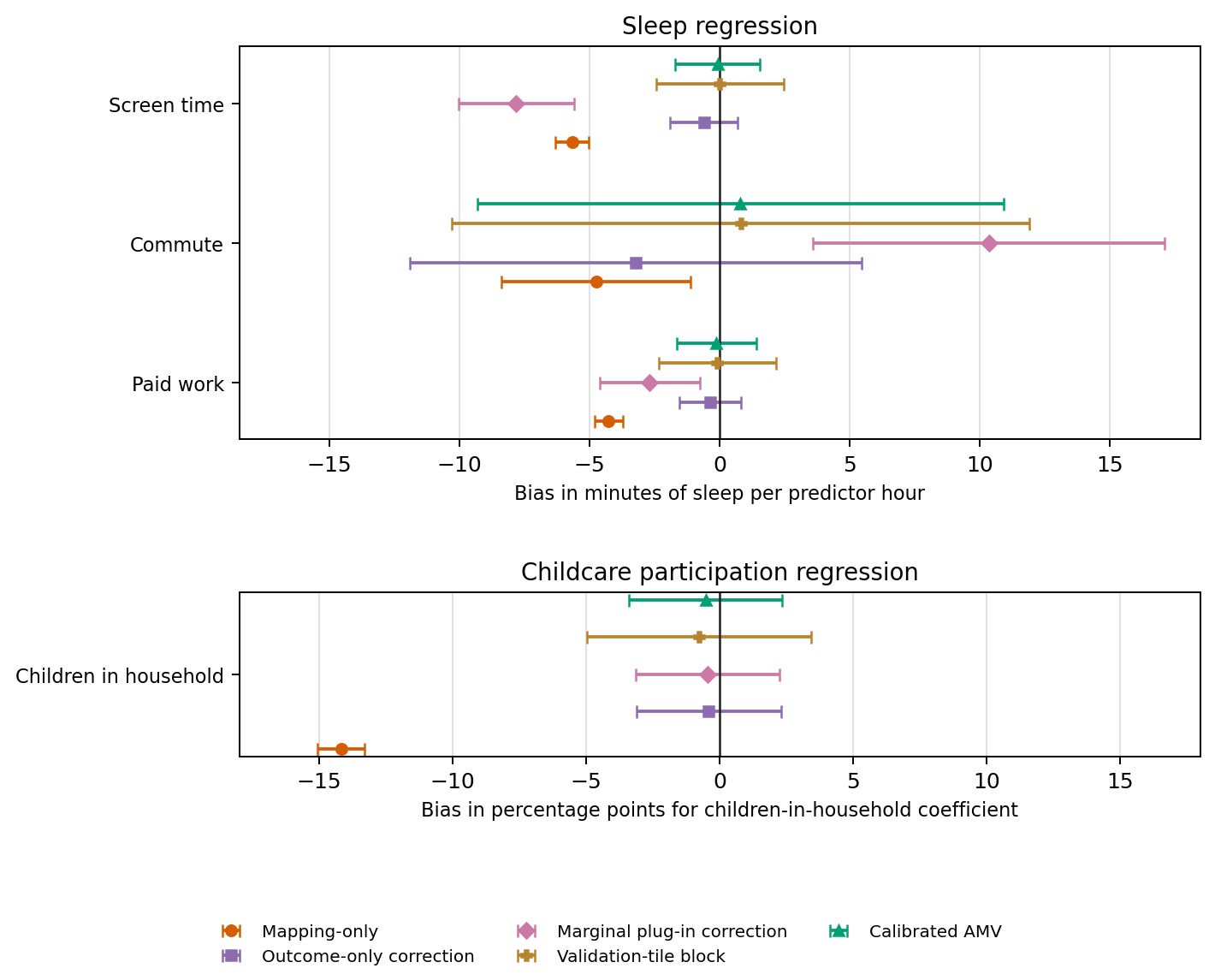}
\caption{ATUS regression correction of biased mapped moments. Points show coefficient bias relative to the complete-diary reference regression target in the moderate-error setting with 18 validation items from the 250-item validation universe. Validation-tile block uses records whose validation tile contains every variable needed for the score. Calibrated AMV uses same-respondent validation to correct the mapped linear-regression moments and tunes the mapped-moment term from other folds. Intervals use the linearized standard-error approximation used to compare methods within this example and do not include full ATUS replicate-weight sampling variance. Sleep coefficients are scaled to minutes of sleep per one-hour predictor change; the childcare coefficient is the percentage-point difference associated with children in the household.}
\label{fig:atus-regression}
%\figalt{Two-panel coefficient-bias plot. In the sleep regression, mapping-only, outcome-only, and marginal plug-in corrections leave visible bias for several predictors, while calibrated AMV is close to zero. The childcare panel shows that mapping-only estimation understates the children-in-household coefficient, while validation-tile block and calibrated AMV are much closer to zero bias.}
\end{figure}

Figure~\ref{fig:atus-regression} reports the regression results in the moderate error setting. For the sleep regression, mapping-only estimation (uncorrected AI-assisted interview) would materially distort all three structured predictor effects, with biases of about \(-4.3\) to \(-5.7\) minutes of sleep per predictor hour.
Correcting only the outcome variable still leaves the commute coefficient biased by about \(-3.2\) minutes, so the validation tiles need to include outcomes and predictors for the same people. Because these two ATUS examples are linear regressions, the implementation uses the linear-moment version in Appendix~\ref{app:linear-moment-regressions}. It corrects the mapped \(X'X\) and \(X'Y\) moments and then solves the resulting normal equations.
The regression estimator first forms calibrated mapped moments, then tunes how much of the mapped moment term to use. In the main \(B=18\) setting the average \(\lambda\) is about 0.78 for the sleep regression and 0.80 for the childcare regression. In the sleep regression, calibrated AMV reduces the commute bias to about \(0.8\) minutes. The commute coefficient remains the least precise sleep coefficient. Commute minutes are sparse and highly skewed, and the regression needs same-respondent validation of sleep, commute, paid work, and screen time. With \(B=18\), the design has enough same-respondent validation to reduce most of this bias, although the commute interval remains wider than the paid-work and screen-time intervals. The block H\'ajek estimate based only on validation tiles uses only records whose validation tile contains every variable needed for the score. In the childcare regression, mapping-only estimation understates the children-in-household coefficient by about 14 percentage points. Calibrated AMV reduces this bias to about 0.5 percentage points and has a smaller standard-error approximation than the validation-only block estimate. In the moderate \(B=18\) setting, calibrated AMV has lower RMSE and a smaller linearized standard-error approximation than this validation-only benchmark for all four displayed coefficients.
 Appendix Table~\ref{tab:atus-regressions} lists supporting coefficients, biases, RMSE values, and standard-error approximations.
Subgroup results show the same pattern. Across sex, employment, education, children-in-household, and weekday/weekend groups, mapping-only sleep bias ranges from about \(-9.5\) to \(-2.1\) minutes, while calibrated AMV subgroup sleep biases are close to zero. For secondary childcare, interview-coding errors vary much more by group, from roughly 101 minutes below to 75 minutes above the complete-diary reference means. The corresponding calibrated AMV subgroup biases are much smaller. These results matter because a design that performs well on average can still leave large subgroup bias. Appendix Table~\ref{tab:atus-subgroups} reports the supporting subgroup ranges. Table~\ref{tab:atus-design-support} reports the supporting uncertainty and question-selection values.

\section{CHAMPS Narratives and Structured Responses}
\label{sec:champs}

The second example uses data from the Child Health and Mortality Prevention Surveillance (CHAMPS) program~\citep{blau2019champs,taylor2020champs,bassat2023champs,champsdata2025}, which monitors health and delivers public health interventions through sites in low-resource settings. We focus on verbal autopsies of children under 5. These are interviews with a surviving caregiver or relative, often after deaths outside hospitals, used to understand the causes and circumstances of death. Field evidence identifies structured-interview length as a practical challenge for verbal-autopsy teams~\citep{surekclark2020va}. That burden makes verbal autopsy a useful setting for asking whether a future conversational protocol could ask fewer structured items while still producing estimates on the structured-response scale.

The CHAMPS data contain de-identified verbal-autopsy and verbal/social-autopsy narratives, translated into English, together with structured verbal-autopsy categorical responses.
The narrative prompt asks the family member for a general description of the circumstances leading to death, with no additional probing. These narratives therefore represent a limited version of an open-response interview. They provide free text that can be translated into structured variables, but they do not include follow-up questions targeted to missing items.  This exercise uses the narratives directly as transcribed.
The narrative is \(D_i\), the structured verbal-autopsy fields define \(Z_i\), and site, age group, sex, and location of death are fully observed case variables in \(A_i\).
\begin{figure}[h!]
\centering
\includegraphics[width=\textwidth]{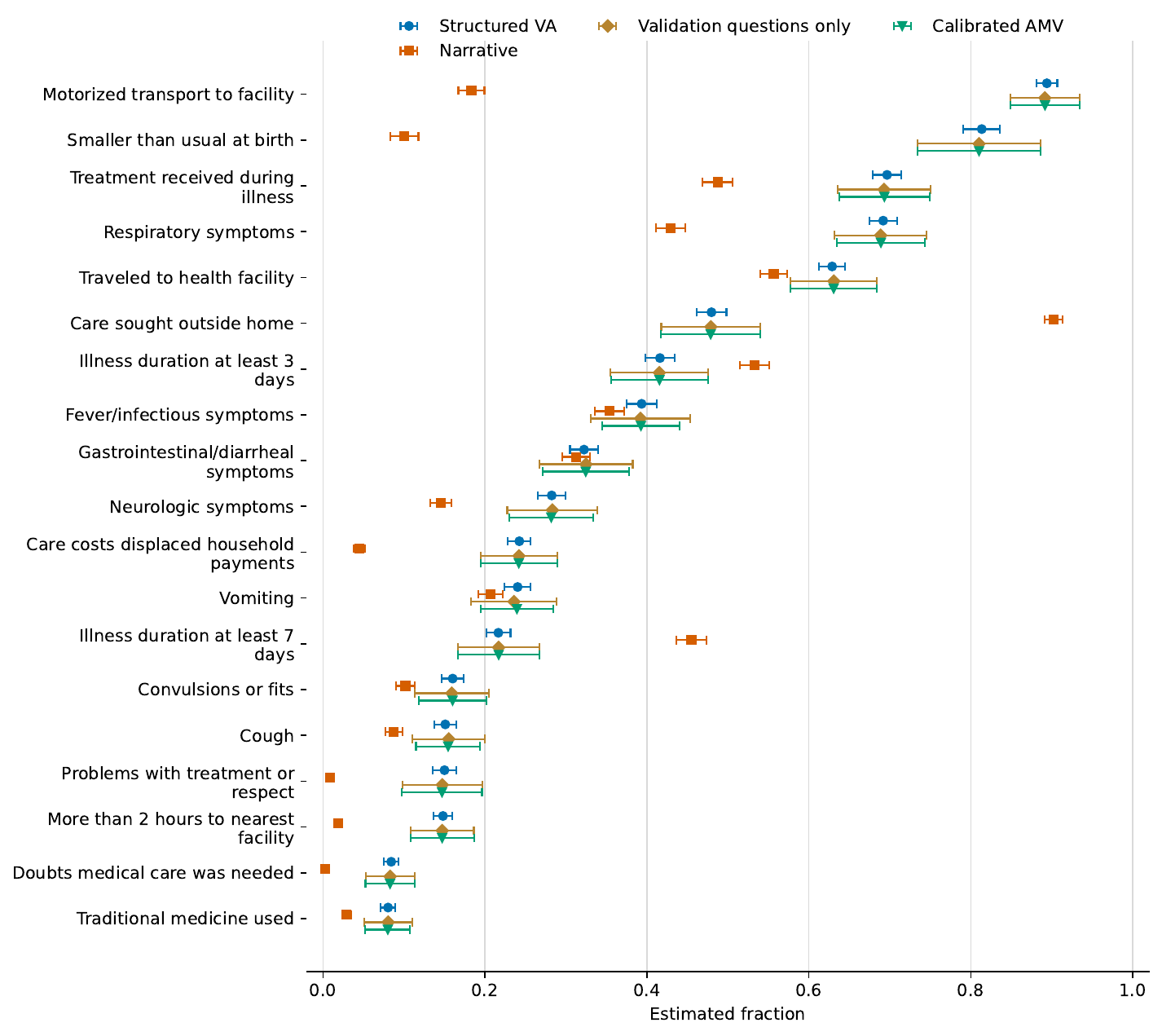}
\caption{Selected CHAMPS structured-response fractions for 19 constructs using existing narratives and AMV. Results use the 4,693 records with nonempty narratives in the 2026-06-01 export. Structured VA is the structured-verbal-autopsy fraction among records with an observed structured response. Narrative uses fixed phrase and duration rules applied to the narrative text, without structured responses or case variables. Validation questions only is the H\'ajek estimate using revealed structured-verbal-autopsy answers. Calibrated AMV uses the Section~\ref{sec:estimators} item estimator in \eqref{eq:item-cal-amv}: the fold-external calibrated mapped value is corrected with revealed structured-verbal-autopsy answers under the same sparse validation rule over 100 validation-assignment draws. Horizontal intervals use stored standard errors where available. Appendix Table~\ref{tab:champs-items-app} reports the numeric values; structured verbal-autopsy responses define the structured-response scale, not clinical truth.}
\label{fig:champs-items}
%\figalt{Horizontal dot-and-interval chart comparing Structured VA, Narrative, Validation questions only, and Calibrated AMV estimates for selected CHAMPS structured-response fractions. Constructs are ordered by the Structured VA fraction.}
\end{figure}
The data contain 9,299 CHAMPS records. Of these, 4,693 have nonempty narratives and 4,606 are blank or missing. Appendix~\ref{app:champs-data} reports the narrative availability, word-length summaries, construct denominators, and data-handling details. This exercise relies entirely on existing narratives as the open component and on structured verbal-autopsy responses from the same respondent as validation items. The ``narrative only'' results use pre-specified phrase and duration rules applied to the narrative text to extract features.
\begin{figure}[h!]
\centering
\includegraphics[width=\textwidth]{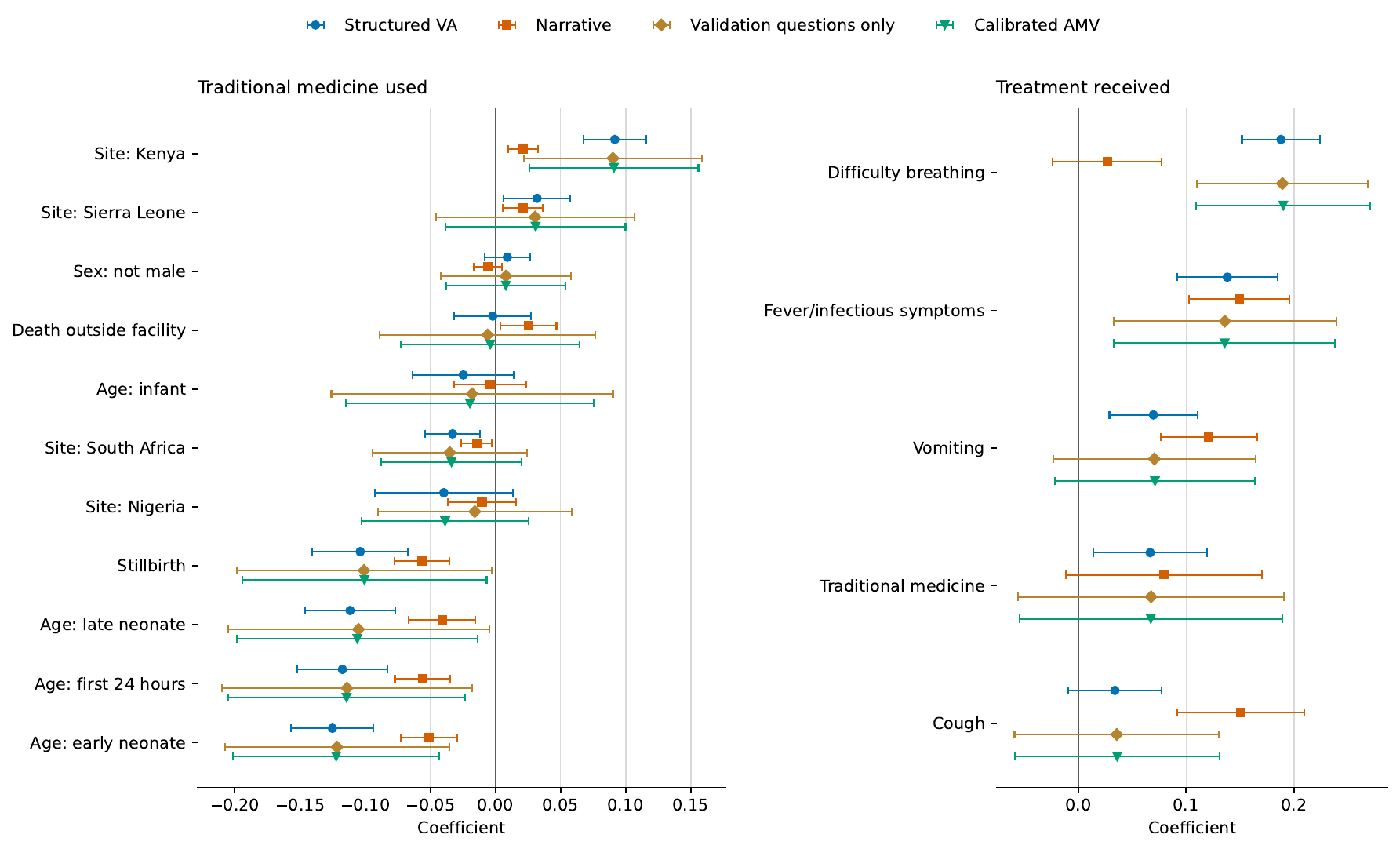}
\caption{Selected CHAMPS regression estimates under same-respondent validation sets. Results use the 4,693 records with nonempty narratives in the 2026-06-01 export. The traditional-medicine model regresses traditional medicine use on age group, site, sex, and whether the death occurred outside a facility. The treatment-received model regresses treatment received during illness on fever/infectious symptoms, cough, difficulty breathing, vomiting, traditional medicine use, age group, and site; the figure displays the five mapped predictors. Narrative uses fixed phrase and duration rules applied to the narrative text, without validation items. Validation questions only solves the score equation using records with the revealed same-respondent validation block, and calibrated AMV solves the calibrated score equation in \eqref{eq:score-cal-amv} with the scalar fold-external score calibration from Section~\ref{sec:estimators}. Structured VA and Narrative intervals use heteroskedasticity-robust linear-model standard errors. Validation questions only and calibrated AMV intervals show variation across 400 validation-assignment draws, so their widths should not be compared with those intervals. Coefficients are descriptive associations on the structured-verbal-autopsy response scale, not causal effects.}
\label{fig:champs-regression}
%\figalt{Two-panel horizontal coefficient plot comparing Structured VA, Narrative, Validation questions only, and Calibrated AMV. The left panel shows a traditional-medicine regression using case-variable predictors. The right panel shows a treatment-received regression using mapped symptom and traditional-medicine predictors. Validation questions only and calibrated AMV are near the Structured VA coefficients, while several Narrative-only coefficients differ sharply.}
\end{figure}

The validation rule uses a 30-construct CHAMPS dictionary as the question universe. Individual items are revealed with probability 0.09. For each displayed regression, the validation-block draw is scaled so that about 9 percent of nonempty-narrative records receive the whole set of validation questions needed for that regression. Across the displayed item fractions, the realized reveal fractions average 0.09. The two regression examples reveal the needed validation questions for about 424 and 421 records on average, or 9.0 percent of the 4,693 nonempty-narrative records. The item results use 100 assignment draws, and the regression results use 400 draws.

Figure~\ref{fig:champs-items} gives the structured-response fraction results, ordered by the structured VA fraction. Because the narratives were collected from a broad prompt rather than from item-specific probes, the fixed narrative rules work unevenly across items. Narrative-only errors are large for several shown items, including motorized transport, smaller than usual at birth, care sought outside home, and treatment received during illness. The largest item-level precision gains occur for symptoms that are often stated in the narrative. For fever, vomiting, and cough, calibrated AMV standard errors are lower than validation-question-only standard errors by about 22 percent, 15 percent, and 12 percent. Smaller gains appear for convulsions, traditional medicine, neurologic symptoms, gastrointestinal symptoms, and respiratory symptoms. Appendix Table~\ref{tab:champs-items-app} gives the supporting numerical comparisons. For the selected items shown here, calibrated AMV under the sparse validation rule is close to the corresponding structured-verbal-autopsy fractions.
Additional numeric values are reported in Appendix Table~\ref{tab:champs-items-app}, while Appendix Table~\ref{tab:champs-goldilocks-mentions} reports direct mention rates for six indicators that are often absent from the open narrative.

Figure~\ref{fig:champs-regression} reports two linear-probability regressions. The first uses traditional medicine as the structured outcome and age group, site, sex, and death location as observed case variables. This example is useful because traditional medicine is sometimes stated directly in the narrative, while the predictors do not require mapping from text. Calibrated AMV is close to the structured-verbal-autopsy coefficients and reduces the average assignment spread by about 9 percent relative to validation questions only; the corresponding RMSE ratio is 0.90.

The second regression uses treatment received during illness as the structured outcome. The five mapped predictors are fever or infectious symptoms, cough, difficulty breathing, vomiting, and traditional medicine use, with age group and site included as controls. These quantities are common in verbal-autopsy and child-mortality work because they describe illness severity, symptom presentation, and care-seeking pathways~\citep{who2024va,li2023openva}. In this more demanding model, the narrative only coefficients differ visibly from the structured-verbal-autopsy coefficients for several predictors, especially difficulty breathing and cough. Calibrated AMV and validation questions only both fall near the structured-verbal-autopsy coefficients. Their across-assignment spreads are nearly identical, with a calibrated-to-validation spread ratio of 0.999. Together, the panels show that narratives help for some quantities and add little for others. The validation questions keep both examples on the structured-response scale. Calibrated AMV improves precision only when the narrative rules add information beyond those validation questions. Appendix Tables~\ref{tab:champs-regression-app} and \ref{tab:champs-regression-support} give the numerical coefficient and support summaries.

\section{Discussion}
\label{sec:discussion}
AMV treats AI-assisted interviewing as a survey measurement design problem, not as ready-to-analyze conversational data. It pairs a conversational interview with sparse, known-probability validation items, then uses those items to estimate means, subgroup quantities, and regressions on the structured-response scale while showing how many structured validation questions are needed. The simulations and ATUS emulation show the burden-precision tradeoff under stated interview-coding error settings. The CHAMPS analysis shows that existing narratives omit many verbal-autopsy items, and that validation items can return selected item and regression estimates to the structured verbal-autopsy scale. Several limits remain.

Spoken dialogue must be transcribed, and often translated across languages, before it can be mapped to structured responses. Multilingual speech-recognition systems such as Whisper perform unevenly across languages; recent Bangla evaluations find that language-adapted alternatives can outperform Whisper on word and character error rates \citep{radford2022robust,ridoy2025adaptability}. Work on language-weighted training, language-model adaptation, and Bangla-specific speech resources is ongoing \citep{pineiromartin2024weighted,dezuazo2025whisperlm,rakib2023oodspeech}, but dialect, code-switching, recording quality, and regional vocabulary can still create group differences in what is captured and coded \citep{koenecke2020racial}.

AMV also addresses only part of what AI-assisted interviews can change. It validates and corrects the structured variables produced from the interview record, but phrasing, probing, perceived privacy, trust in automated interviewers, cultural expectations about AI use, device quality, connectivity, literacy, disability, and comfort with conversational tools can affect who responds, what they disclose, and how they interpret the exchange \citep{tourangeau2000psychology,schober1997conversational,conrad2000clarifying,kreuter2025aapor,aapor2026responsibleai}. These issues need direct study because they can create group differences before the validation design is applied.

Finally, there are substantial opportunities for next steps.  This paper treats unstructured data as spoken dialogue, but the same logic could extend to other respondent-provided material: photos of meals, product labels or shelf prices, receipts, housing conditions, or short videos recorded during an AI-assisted interview \citep{shao2021integrated,tahir2021comprehensive,wireduk2016premise}. Those settings could combine automated extraction with validation items or human review on a known-probability subset, but image and video data would also require methods for objects, locations, timestamps, repeated frames, privacy masking, and correlated errors within each uploaded file.

Overall, this paper articulates the potential and caveats associated with AI-assisted interviews.  AI-assisted interviewing will not automatically reduce respondent burden. It will reduce burden only when the AI maps responses well enough that a small number of
  validation questions can correct what remains. The formulas presented here let practitioners check whether that threshold is met before data collection begins. When it is,
  the efficiency gains are real. When it is not, the same framework says so clearly and points to the remedy, which is to ask more validation questions, probe more directly, or
  measure the item with a structured question.

\bibliographystyle{apalike}
\bibliography{references}

@article{groves2006responsive,
  author = {Groves, Robert M. and Heeringa, Steven G.},
  title = {Responsive Design for Household Surveys: Tools for Actively Controlling Survey Errors and Costs},
  journal = {Journal of the Royal Statistical Society: Series A (Statistics in Society)},
  year = {2006},
  volume = {169},
  number = {3},
  pages = {439--457},
  doi = {10.1111/j.1467-985X.2006.00423.x}
}

@misc{BLS2025ATUSMicrodata,
  author       = {{U.S. Bureau of Labor Statistics}},
  title        = {American Time Use Survey: ATUS 2003--2024 Multi-Year Microdata Files},
  year         = {2025},
  url          = {https://www.bls.gov/tus/data/datafiles-0324.htm},
  note         = {Last modified 2025-06-26; accessed 2026-06-02}
}

@misc{CensusACS2025Questionnaire,
  author       = {{U.S. Census Bureau}},
  title        = {American Community Survey: Informational Copy},
  year         = {2025},
  url          = {https://www2.census.gov/programs-surveys/acs/methodology/questionnaires/2025/quest25.pdf},
  note         = {Form ACS-1(INFO)(2025)}
}

@article{HamermeshFrazisStewart2005ATUS,
  author       = {Hamermesh, Daniel S. and Frazis, Harley and Stewart, Jay},
  title        = {Data Watch: The American Time Use Survey},
  journal      = {Journal of Economic Perspectives},
  year         = {2005},
  volume       = {19},
  number       = {1},
  pages        = {221--232},
  doi          = {10.1257/0895330053148029}
}

@article{robins1994estimation,
  author = {Robins, James M. and Rotnitzky, Andrea and Zhao, Lue Ping},
  title = {Estimation of Regression Coefficients When Some Regressors Are Not Always Observed},
  journal = {Journal of the American Statistical Association},
  year = {1994},
  volume = {89},
  number = {427},
  pages = {846--866},
  doi = {10.1080/01621459.1994.10476818}
}

@article{bang2005doubly,
  author = {Bang, Heejung and Robins, James M.},
  title = {Doubly Robust Estimation in Missing Data and Causal Inference Models},
  journal = {Biometrics},
  year = {2005},
  volume = {61},
  number = {4},
  pages = {962--973},
  doi = {10.1111/j.1541-0420.2005.00377.x}
}

@book{tsiatis2006semiparametric,
  author = {Tsiatis, Anastasios A.},
  title = {Semiparametric Theory and Missing Data},
  publisher = {Springer},
  year = {2006}
}

@article{radford2022robust,
  author = {Radford, Alec and Kim, Jong Wook and Xu, Tao and Brockman, Greg and McLeavey, Christine and Sutskever, Ilya},
  title = {Robust Speech Recognition via Large-Scale Weak Supervision},
  journal = {arXiv preprint},
  year = {2022},
  eprint = {2212.04356},
  archivePrefix = {arXiv},
  url = {https://arxiv.org/abs/2212.04356}
}

@article{ridoy2025adaptability,
  author = {Ridoy, Md Sazzadul Islam and Akter, Sumi and Rahman, Md. Aminur},
  title = {Adaptability of {ASR} Models on Low-Resource Language: A Comparative Study of {Whisper} and {Wav2Vec-BERT} on {Bangla}},
  journal = {arXiv preprint},
  year = {2025},
  eprint = {2507.01931},
  archivePrefix = {arXiv},
  url = {https://arxiv.org/abs/2507.01931}
}

@article{pineiromartin2024weighted,
  author = {Pi{\~n}eiro-Mart{\'i}n, Andr{\'e}s and Garc{\'i}a-Mateo, Carmen and Doc{\'i}o-Fern{\'a}ndez, Laura and L{\'o}pez-P{\'e}rez, Mar{\'i}a del Carmen and Rehm, Georg},
  title = {Weighted Cross-entropy for Low-Resource Languages in Multilingual Speech Recognition},
  journal = {arXiv preprint},
  year = {2024},
  eprint = {2409.16954},
  archivePrefix = {arXiv},
  url = {https://arxiv.org/abs/2409.16954}
}

@article{dezuazo2025whisperlm,
  author = {de Zuazo, Xabier and Navas, Eva and Saratxaga, Ibon and Rioja, Inma Hern{\'a}ez},
  title = {{Whisper-LM}: Improving {ASR} Models with Language Models for Low-Resource Languages},
  journal = {arXiv preprint},
  year = {2025},
  eprint = {2503.23542},
  archivePrefix = {arXiv},
  url = {https://arxiv.org/abs/2503.23542}
}

@article{rakib2023oodspeech,
  author = {Rakib, Fazle Rabbi and Dip, Souhardya Saha and Alam, Samiul and Tasnim, Nazia and Shihab, Md. Istiak Hossain and Ansary, Md. Nazmuddoha and Hossen, Syed Mobassir and Meghla, Marsia Haque and Mamun, Mamunur and Sadeque, Farig and Chowdhury, Sayma Sultana and Reasat, Tahsin and Sushmit, Asif and Humayun, Ahmed Imtiaz},
  title = {{OOD-Speech}: A Large {Bengali} Speech Recognition Dataset for Out-of-Distribution Benchmarking},
  journal = {arXiv preprint},
  year = {2023},
  eprint = {2305.09688},
  archivePrefix = {arXiv},
  url = {https://arxiv.org/abs/2305.09688}
}

@article{koenecke2020racial,
  author = {Koenecke, Allison and Nam, Andrew and Lake, Emily and Nudell, Joe and Quartey, Minnie and Mengesha, Zion and Toups, Connor and Rickford, John R. and Jurafsky, Dan and Goel, Sharad},
  title = {Racial Disparities in Automated Speech Recognition},
  journal = {Proceedings of the National Academy of Sciences},
  year = {2020},
  volume = {117},
  number = {14},
  pages = {7684--7689},
  doi = {10.1073/pnas.1915768117}
}

@article{shao2021integrated,
  author = {Shao, Zeman and Han, Yue and He, Jiangpeng and Mao, Runyu and Wright, Janine and Kerr, Deborah and Boushey, Carol and Zhu, Fengqing},
  title = {An Integrated System for Mobile Image-Based Dietary Assessment},
  journal = {arXiv preprint},
  year = {2021},
  eprint = {2110.01754},
  archivePrefix = {arXiv},
  url = {https://arxiv.org/abs/2110.01754}
}

@article{tahir2021comprehensive,
  author = {Tahir, Ghalib and Loo, Chu Kiong},
  title = {A Comprehensive Survey of Image-Based Food Recognition and Volume Estimation Methods for Dietary Assessment},
  journal = {arXiv preprint},
  year = {2021},
  eprint = {2106.11776},
  archivePrefix = {arXiv},
  url = {https://arxiv.org/abs/2106.11776}
}

@misc{wireduk2016premise,
  author = {Baker, David},
  title = {Photos Are Creating a Real-Time Food-Price Index},
  howpublished = {WIRED UK},
  year = {2016},
  month = apr,
  url = {https://www.wired.com/story/premise-app-food-tracking-brazil-philippines/}
}

@techreport{barari2024generative,
  author = {Barari, Soubhik and Slowinski, Zoe and Wang, Natalie and Angbazo, Jarret and Sepulvado, Brandon and Christian, Leah and Dean, Elizabeth},
  title = {Generative {AI} Can Enhance Survey Interviews},
  institution = {NORC at the University of Chicago},
  type = {Research Brief},
  year = {2024},
  month = nov,
  url = {https://www.norc.org/research/library/generative-ai-can-enhance-survey-interviews.html}
}

@article{xiao2020tell,
  author = {Xiao, Ziang and Zhou, Michelle X. and Liao, Q. Vera and Mark, Gloria and Chi, Changyan and Chen, Wenxi and Yang, Huahai},
  title = {Tell Me About Yourself: Using an {AI}-Powered Chatbot to Conduct Conversational Surveys with Open-Ended Questions},
  journal = {ACM Transactions on Computer-Human Interaction},
  year = {2020},
  volume = {27},
  number = {3},
  articleno = {15},
  pages = {1--37},
  doi = {10.1145/3381804}
}

@inproceedings{wuttke2025ai,
  author = {Wuttke, Alexander and A{\ss}enmacher, Matthias and Klamm, Christopher and Lang, Max M. and W{\"u}rschinger, Quirin and Kreuter, Frauke},
  title = {{AI} Conversational Interviewing: Transforming Surveys with {LLM}s as Adaptive Interviewers},
  booktitle = {Proceedings of the 9th Joint SIGHUM Workshop on Computational Linguistics for Cultural Heritage, Social Sciences, Humanities and Literature},
  year = {2025},
  pages = {179--204},
  address = {Albuquerque, New Mexico},
  publisher = {Association for Computational Linguistics},
  doi = {10.18653/v1/2025.latechclfl-1.17},
  url = {https://aclanthology.org/2025.latechclfl-1.17/}
}

@article{barari2025ai,
  author = {Barari, Soubhik and Angbazo, Jarret and Wang, Natalie and Christian, Leah M. and Dean, Elizabeth and Slowinski, Zoe and Sepulvado, Brandon},
  title = {{AI}-Assisted Conversational Interviewing: Effects on Data Quality and Respondent Experience},
  journal = {arXiv preprint},
  year = {2025},
  eprint = {2504.13908},
  archivePrefix = {arXiv},
  doi = {10.48550/arXiv.2504.13908},
  url = {https://arxiv.org/abs/2504.13908}
}

@article{groves2010total,
  author = {Groves, Robert M. and Lyberg, Lars},
  title = {Total Survey Error: Past, Present, and Future},
  journal = {Public Opinion Quarterly},
  year = {2010},
  volume = {74},
  number = {5},
  pages = {849--879},
  doi = {10.1093/poq/nfq065}
}

@article{biemer2010total,
  author = {Biemer, Paul P.},
  title = {Total Survey Error: Design, Implementation, and Evaluation},
  journal = {Public Opinion Quarterly},
  year = {2010},
  volume = {74},
  number = {5},
  pages = {817--848},
  doi = {10.1093/poq/nfq058}
}

@article{tourangeau2017adaptive,
  author = {Tourangeau, Roger and Brick, J. Michael and Lohr, Sharon and Li, Jane},
  title = {Adaptive and Responsive Survey Designs: A Review and Assessment},
  journal = {Journal of the Royal Statistical Society: Series A (Statistics in Society)},
  year = {2017},
  volume = {180},
  number = {1},
  pages = {203--223},
  doi = {10.1111/rssa.12186}
}

@article{schouten2009indicators,
  author = {Schouten, Barry and Cobben, Fannie and Bethlehem, Jelke},
  title = {Indicators for the Representativeness of Survey Response},
  journal = {Survey Methodology},
  year = {2009},
  volume = {35},
  number = {1},
  pages = {101--113},
  url = {https://www150.statcan.gc.ca/n1/pub/12-001-x/2009001/article/10887-eng.pdf}
}

@article{wagner2012alternative,
  author = {Wagner, James},
  title = {A Comparison of Alternative Indicators for the Risk of Nonresponse Bias},
  journal = {Public Opinion Quarterly},
  year = {2012},
  volume = {76},
  number = {3},
  pages = {555--575},
  doi = {10.1093/poq/nfs032}
}

@article{raghunathan1995split,
  author = {Raghunathan, Trivellore E. and Grizzle, James E.},
  title = {A Split Questionnaire Survey Design},
  journal = {Journal of the American Statistical Association},
  year = {1995},
  volume = {90},
  number = {429},
  pages = {54--63},
  doi = {10.1080/01621459.1995.10476488}
}

@article{graham2006planned,
  title={Planned missing data designs in psychological research.},
  author={Graham, John W and Taylor, Bonnie J and Olchowski, Allison E and Cumsille, Patricio E},
  journal={Psychological methods},
  volume={11},
  number={4},
  pages={323-343},
  year={2006},
  publisher={American Psychological Association}
}

@inproceedings{gonzalez2008multiplematrix,
  title={Multiple matrix sampling: A review},
  author={Gonzalez, Jeffrey M and Eltinge, John L},
  booktitle={Proceedings of the Section on Survey Research Methods, American Statistical Association},
  pages={3069--3075},
  year={2007},
  organization={American Statistical Association Alexandria, VA}
}

@article{thomas2006matrix,
  title={An evaluation of matrix sampling methods using data from the National Health and Nutrition Examination Survey},
  author={Thomas, Neal and Raghunathan, Trivellore E and Schenker, Nathaniel and Katzoff, Myron J and Johnson, Clifford L},
  journal={Survey Methodology},
  volume={32},
  number={2},
  pages={217},
  year={2006}
}

@article{axenfeld2022split,
  author = {Axenfeld, Julian B. and Blom, Annelies G. and Bruch, Christian and Wolf, Christof},
  title = {Split Questionnaire Designs for Online Surveys: The Impact of Module Construction on Imputation Quality},
  journal = {Journal of Survey Statistics and Methodology},
  year = {2022},
  volume = {10},
  number = {5},
  pages = {1236--1262},
  doi = {10.1093/jssam/smab055}
}

@article{montgomery2013cat,
  author = {Montgomery, Jacob M. and Cutler, Josh},
  title = {Computerized Adaptive Testing for Public Opinion Surveys},
  journal = {Political Analysis},
  year = {2013},
  volume = {21},
  number = {2},
  pages = {172--192},
  doi = {10.1093/pan/mps060}
}

@article{zhang2020active,
  title={Active matrix factorization for surveys},
  author={Zhang, Chelsea and Taylor, Sean J and Cobb, Curtiss and Sekhon, Jasjeet},
  journal={The Annals of Applied Statistics},
  volume={14},
  number={3},
  pages={1182--1206},
  year={2020},
  publisher={JSTOR}
}

@inproceedings{yoshida2023bayesian,
  author = {Yoshida, Toshiya and Fan, Trinity Shuxian and McCormick, Tyler H. and Wu, Zhenke and Li, Zehang Richard},
  title = {Bayesian Active Questionnaire Design for Cause-of-Death Assignment Using Verbal Autopsies},
  booktitle = {Proceedings of the Conference on Health, Inference, and Learning},
  series = {Proceedings of Machine Learning Research},
  year = {2023},
  volume = {209},
  pages = {37--49},
  publisher = {PMLR},
  url = {https://proceedings.mlr.press/v209/yoshida23a.html}
}

@techreport{aapor2026responsibleai,
  author = {{AAPOR Task Force on Responsible AI Integration in Survey Research}},
  title = {Responsible {AI} Integration in Survey Research},
  institution = {American Association for Public Opinion Research},
  year = {2026},
  url = {https://aapor.org/wp-content/uploads/2026/05/Responsible-AI-Integration-In-Survey-Research.pdf}
}

@article{halterman2025codebook,
  author = {Halterman, Andrew and Keith, Katherine},
  title = {Codebook {LLM}s: Evaluating {LLM}s as Measurement Tools for Political Science Concepts},
  journal = {Political Analysis},
  year = {2025},
  doi = {10.1017/pan.2025.10017},
  url = {https://www.cambridge.org/core/journals/political-analysis/article/codebook-llms-evaluating-llms-as-measurement-tools-for-political-science-concepts/7B323A0E47F782F2698A0AE849EA00DE}
}

@article{mellon2024issues,
  title={Do {AI}s know what the most important issue is? Using language models to code open-text social survey responses at scale},
  author={Mellon, Jonathan and Bailey, Jack and Scott, Ralph and Breckwoldt, James and Miori, Marta and Schmedeman, Phillip},
  journal={Research \& Politics},
  volume={11},
  number={1},
  pages={20531680241231468},
  year={2024},
  publisher={SAGE Publications Sage UK: London, England}
}

@article{digiuseppe2026paired,
  author = {DiGiuseppe, Matthew R. and Flynn, Michael E.},
  title = {Scaling Open-Ended Survey Responses Using {LLM}-Paired Comparisons},
  journal = {Public Opinion Quarterly},
  year = {2026},
  doi = {10.1093/poq/nfag013},
  url = {https://academic.oup.com/poq/advance-article/doi/10.1093/poq/nfag013/8551356}
}

@article{schober1997conversational,
  author = {Schober, Michael F. and Conrad, Frederick G.},
  title = {Does Conversational Interviewing Reduce Survey Measurement Error?},
  journal = {Public Opinion Quarterly},
  year = {1997},
  volume = {61},
  number = {4},
  pages = {576--602},
  doi = {10.1086/297818}
}

@article{conrad2000clarifying,
  author = {Conrad, Frederick G. and Schober, Michael F.},
  title = {Clarifying Question Meaning in a Household Telephone Survey},
  journal = {Public Opinion Quarterly},
  year = {2000},
  volume = {64},
  number = {1},
  pages = {1--28},
  doi = {10.1086/316757}
}

@article{frisbie1968computers,
  author = {Frisbie, Bruce and Sudman, Seymour},
  title = {The Use of Computers in Coding Free Responses},
  journal = {Public Opinion Quarterly},
  year = {1968},
  volume = {32},
  number = {2},
  pages = {216--232},
  doi = {10.1086/267600}
}

@article{kreuter2025aapor,
  author = {Kreuter, Frauke},
  title = {{AAPOR} Presidential Address: ``That Ain't the Way I Heard It!'' On the Role of Surveys in Shaping (and Being Shaped by) Generative {AI}},
  journal = {Public Opinion Quarterly},
  year = {2025},
  volume = {89},
  number = {3},
  pages = {950--962},
  doi = {10.1093/poq/nfaf049}
}

@article{han2025word,
  author = {Han, Ze and Truex, Rory and Liu, Naijia},
  title = {Measuring Political Attitudes with Word Association},
  journal = {Public Opinion Quarterly},
  year = {2025},
  volume = {89},
  number = {4},
  pages = {965--997},
  doi = {10.1093/poq/nfaf038}
}

@article{landesvatter2026speech,
  author = {Landesvatter, Camille and Behnert, Jan and Bauer, Paul C.},
  title = {Comparing Speech-to-Text Algorithms for Transcribing Voice Data from Surveys},
  journal = {Public Opinion Quarterly},
  year = {2026},
  volume = {89},
  number = {4},
  pages = {1154--1166},
  doi = {10.1093/poq/nfaf056}
}

@book{tourangeau2000psychology,
  author = {Tourangeau, Roger and Rips, Lance J. and Rasinski, Kenneth},
  title = {The Psychology of Survey Response},
  publisher = {Cambridge University Press},
  address = {Cambridge},
  year = {2000}
}

@book{cochran1977sampling,
  author = {Cochran, William G.},
  title = {Sampling Techniques},
  edition = {3},
  publisher = {Wiley},
  address = {New York},
  year = {1977}
}

@article{chen2000unified,
  author = {Chen, Y. H. and Chen, H.},
  title = {A Unified Approach to Regression Analysis under Double-Sampling Designs},
  journal = {Journal of the Royal Statistical Society: Series B (Statistical Methodology)},
  year = {2000},
  volume = {62},
  number = {3},
  pages = {449--460},
  doi = {10.1111/1467-9868.00243}
}

@book{sarndal1992model,
  author = {S{\"a}rndal, Carl-Erik and Swensson, Bengt and Wretman, Jan},
  title = {Model Assisted Survey Sampling},
  publisher = {Springer},
  address = {New York},
  year = {1992}
}

@article{angelopoulos2023prediction,
  author = {Angelopoulos, Anastasios N. and Bates, Stephen and Fannjiang, Clara and Jordan, Michael I. and Zrnic, Tijana},
  title = {Prediction-Powered Inference},
  journal = {Science},
  year = {2023},
  volume = {382},
  number = {6671},
  pages = {669--674},
  doi = {10.1126/science.adi6000}
}

@article{zrnic2024cross,
  author = {Zrnic, Tijana and Cand{\`e}s, Emmanuel J.},
  title = {Cross-Prediction-Powered Inference},
  journal = {Proceedings of the National Academy of Sciences},
  year = {2024},
  volume = {121},
  number = {15},
  pages = {e2322083121},
  doi = {10.1073/pnas.2322083121}
}

@article{wang2020methods,
  author = {Wang, Siruo and McCormick, Tyler H. and Leek, Jeffrey T.},
  title = {Methods for Correcting Inference Based on Outcomes Predicted by Machine Learning},
  journal = {Proceedings of the National Academy of Sciences},
  year = {2020},
  volume = {117},
  number = {48},
  pages = {30266--30275},
  doi = {10.1073/pnas.2001238117}
}

@article{salerno2025modern,
  author = {Salerno, Stephen and Hoffman, Kentaro and Afiaz, Awan and Neufeld, Anna and McCormick, Tyler H. and Leek, Jeffrey T.},
  title = {Do We Really Even Need Data? A Modern Look at Drawing Inference with Predicted Data},
  journal = {arXiv preprint},
  year = {2025},
  eprint = {2512.05456},
  archivePrefix = {arXiv},
  doi = {10.48550/arXiv.2512.05456},
  url = {https://arxiv.org/abs/2512.05456}
}

@article{salerno2025moment,
  author = {Salerno, Stephen and Hoffman, Kentaro and Afiaz, Awan and Neufeld, Anna and McCormick, Tyler H. and Leek, Jeffrey T.},
  title = {A Moment-Based Generalization to Post-Prediction Inference},
  journal = {arXiv preprint},
  year = {2025},
  eprint = {2507.09119},
  archivePrefix = {arXiv},
  doi = {10.48550/arXiv.2507.09119},
  url = {https://arxiv.org/abs/2507.09119}
}

@article{salerno2026spatial,
  author = {Salerno, Stephen and Wu, Zhenke and McCormick, Tyler H.},
  title = {Spatially Robust Inference with Predicted and Missing at Random Labels},
  journal = {arXiv preprint},
  year = {2026},
  eprint = {2603.11368},
  archivePrefix = {arXiv},
  doi = {10.48550/arXiv.2603.11368},
  url = {https://arxiv.org/abs/2603.11368}
}

@article{visokay2025obesity,
  author = {Visokay, A. and Hoffman, K. and Salerno, S. and McCormick, T. H. and Johfre, S.},
  title = {How to Measure Obesity in Public Health Research? Problems with Using {BMI} for Population Inference},
  journal = {medRxiv preprint},
  year = {2025},
  doi = {10.1101/2025.04.01.25325037},
  url = {https://www.medrxiv.org/content/10.1101/2025.04.01.25325037v1}
}

@inproceedings{fan2024narratives,
  author = {Fan, Shuxian and Visokay, Adam and Hoffman, Kentaro and Salerno, Stephen and Liu, Li and Leek, Jeffrey T. and McCormick, Tyler H.},
  title = {From Narratives to Numbers: Valid Inference Using Language Model Predictions from Verbal Autopsies},
  booktitle = {Proceedings of the Conference on Language Modeling},
  year = {2024},
  eprint = {2404.02438},
  archivePrefix = {arXiv},
  doi = {10.48550/arXiv.2404.02438},
  url = {https://openreview.net/forum?id=QbCHlIqbDJ}
}

@misc{egami2023imperfect,
  author = {Egami, Naoki and Hinck, Musashi and Stewart, Brandon M. and Wei, Hanying},
  title = {Using Imperfect Surrogates for Downstream Inference: Design-Based Supervised Learning for Social Science Applications of Large Language Models},
  year = {2023},
  eprint = {2306.04746},
  archivePrefix = {arXiv},
  doi = {10.48550/arXiv.2306.04746},
  url = {https://arxiv.org/abs/2306.04746}
}

@article{blau2019champs,
  title={Overview and development of the child health and mortality prevention surveillance determination of cause of death (DeCoDe) process and DeCoDe diagnosis standards},
  author={Blau, Dianna M and Caneer, J Patrick and Philipsborn, Rebecca P and Madhi, Shabir A and Bassat, Quique and Varo, Rosauro and Mandomando, In{\'a}cio and Igunza, Kitiezo Aggrey and Kotloff, Karen L and Tapia, Milagritos D and others},
  journal={Clinical Infectious Diseases},
  volume={69},
  number={Supplement\_4},
  pages={S333--S341},
  year={2019},
  publisher={Oxford University Press US}
}

@article{taylor2020champs,
  title={Initial findings from a novel population-based child mortality surveillance approach: a descriptive study},
  author={Taylor, Allan W and Blau, Dianna M and Bassat, Quique and Onyango, Dickens and Kotloff, Karen L and El Arifeen, Shams and Mandomando, Inacio and Chawana, Richard and Baillie, Vicky L and Akelo, Victor and others},
  journal={The Lancet Global Health},
  volume={8},
  number={7},
  pages={e909--e919},
  year={2020},
  publisher={Elsevier}
}

@incollection{surekclark2020va,
  author = {Surek-Clark, Clarissa},
  title = {Verbal Autopsy Interview Standardization Study: A Report from the Field},
  booktitle = {The Anthropological Demography of Health},
  editor = {Petit, V{\'e}ronique and Qureshi, Kaveri and Charbit, Yves and Kreager, Philip},
  publisher = {Oxford University Press},
  year = {2020},
  pages = {301--320},
  doi = {10.1093/oso/9780198862437.003.0011}
}

@misc{champsdata2025,
  author = {{Child Health and Mortality Prevention Surveillance Network}},
  title = {{CHAMPS} Data Downloads and Data Structure Documentation},
  year = {2025},
  url = {https://champshealth.org/data/}
}

@misc{who2024va,
  author = {{World Health Organization}},
  title = {Verbal Autopsy Standards: Ascertaining and Attributing Causes of Death Tool},
  year = {2024},
  url = {https://www.who.int/standards/classifications/other-classifications/verbal-autopsy-standards-ascertaining-and-attributing-causes-of-death-tool}
}

@article{li2023openva,
  author = {Li, Zehang Richard and Thomas, Jason and Choi, Eungang and McCormick, Tyler H. and Clark, Samuel J.},
  title = {The openVA Toolkit for Verbal Autopsies},
  journal = {The R Journal},
  year = {2023},
  volume = {14},
  number = {4},
  pages = {316--334},
  doi = {10.32614/RJ-2023-020},
  url = {https://journal.r-project.org/articles/RJ-2023-020/}
}

@article{basner2007sleep,
  author = {Basner, Mathias and Fomberstein, Kenneth M. and Razavi, Farid M. and Banks, Siobhan and William, Jeffrey H. and Rosa, Roger R. and Dinges, David F.},
  title = {American Time Use Survey: Sleep Time and Its Relationship to Waking Activities},
  journal = {Sleep},
  year = {2007},
  volume = {30},
  number = {9},
  pages = {1085--1095},
  doi = {10.1093/sleep/30.9.1085}
}

@article{guryan2008parental,
  author = {Guryan, Jonathan and Hurst, Erik and Kearney, Melissa},
  title = {Parental Education and Parental Time with Children},
  journal = {Journal of Economic Perspectives},
  year = {2008},
  volume = {22},
  number = {3},
  pages = {23--46},
  doi = {10.1257/jep.22.3.23}
}

@article{raley2012fathers,
  author = {Raley, Sara and Bianchi, Suzanne M. and Wang, Wendy},
  title = {When Do Fathers Care? Mothers' Economic Contribution and Fathers' Involvement in Child Care},
  journal = {American Journal of Sociology},
  year = {2012},
  volume = {117},
  number = {5},
  pages = {1422--1459},
  doi = {10.1086/663354}
}

@article{pepin2018marital,
  author = {Pepin, Joanna R. and Sayer, Liana C. and Casper, Lynne M.},
  title = {Marital Status and Mothers' Time Use: Childcare, Housework, Leisure, and Sleep},
  journal = {Demography},
  year = {2018},
  volume = {55},
  number = {1},
  pages = {107--133},
  doi = {10.1007/s13524-018-0647-x}
}

@article{bassat2023champs,
  author = {Bassat, Quique and others},
  title = {Causes of Death Among Infants and Children in the Child Health and Mortality Prevention Surveillance Network},
  journal = {JAMA Network Open},
  year = {2023},
  volume = {6},
  number = {7},
  pages = {e2322494},
  doi = {10.1001/jamanetworkopen.2023.22494}
}

@article{gronsbell2024anotherlook,
  author = {Gronsbell, Jessica and Gao, Jianhui and McCaw, Zachary R. and Shi, Yaqi and Cheng, David},
  title = {Another Look at Statistical Inference with Machine Learning-Imputed Data},
  journal = {arXiv preprint},
  year = {2024},
  eprint = {2411.19908},
  archivePrefix = {arXiv},
  doi = {10.48550/arXiv.2411.19908},
  url = {https://arxiv.org/abs/2411.19908}
}

@article{chen2025unified,
  author = {Chen, Xingran and McCormick, Tyler H. and Mukherjee, Bhramar and Wu, Zhenke},
  title = {A Unified Framework for Inference with General Missingness Patterns and Machine Learning Imputation},
  journal = {arXiv preprint},
  year = {2025},
  eprint = {2508.15162},
  archivePrefix = {arXiv},
  doi = {10.48550/arXiv.2508.15162},
  url = {https://arxiv.org/abs/2508.15162}
}

@article{bose1943samplingerror,
  author = {Bose, Chameli},
  title = {Note on the Sampling Error in the Method of Double Sampling},
  journal = {Sankhy{\=a}: The Indian Journal of Statistics (1933--1960)},
  year = {1943},
  volume = {6},
  number = {3},
  pages = {329--330},
  url = {http://www.jstor.org/stable/25047787},
  issn = {00364452}
}

\newpage
\appendix
\numberwithin{table}{section}
\numberwithin{figure}{section}

\section{Adjacent Literatures}
\label{app:related-literatures}
We provide a table that gives a conceptual overview of how our work fits into this growing field.

\begin{table}[h!]
\centering
\caption{Adjacent methods and the role they play in this paper}
\label{tab:related}
\footnotesize
\setlength{\tabcolsep}{3pt}
\renewcommand{\arraystretch}{1.05}
\begin{singlespace}
\begin{tabularx}{\textwidth}{p{0.20\textwidth}X X}
\toprule
Literature & Relevance & Boundary for this paper \\
\midrule
Total survey error and survey response process & Provides the broad accounting of representation and measurement errors, and explains why question comprehension, retrieval, judgment, and response mapping matter for data quality \citep{biemer2010total,groves2010total,tourangeau2000psychology}. & Leaves open how to validate a model-mediated interview that both elicits natural language and maps it into structured survey responses. \\
Conversational interviewing & Shows that flexible clarification can improve comprehension and reduce some measurement errors, while changing cost and interaction dynamics \citep{schober1997conversational}. & Studies human interviewing more than versioned AI systems with logged prompts, model versions, probabilistic coding, and validation-question propensities. \\
Responsive and adaptive survey design, including R-indicators & Uses paradata and design changes to manage survey error, cost, and representation during data collection \citep{groves2006responsive,tourangeau2017adaptive,schouten2009indicators,wagner2012alternative}. & Usually adapts contact protocols, modes, incentives, or follow-up effort. The present problem adapts and validates measurement content inside the interview. \\
Split questionnaires, matrix sampling, and planned missingness & Shows that long instruments can be split across respondents and analyzed using design or imputation logic \citep{raghunathan1995split,graham2006planned,gonzalez2008multiplematrix,thomas2006matrix,axenfeld2022split}. & Typically lacks a full mapped structured matrix for all respondents, natural-language evidence, uncertainty flags, and adaptive validation-item selection tied to interview-coding error. \\
Computerized adaptive testing, active matrix factorization, and active verbal autopsy & Selects informative items to reduce burden while preserving measurement or coding performance \citep{montgomery2013cat,zhang2020active,yoshida2023bayesian}. & Often targets individual latent traits, matrix completion, or cause assignment. Our target is population and regression inference for a reusable structured-response matrix. \\
AI conversational interviewing, speech processing, and LLM coding & Evaluates AI probing, speech transcription, LLM coding, scaling, and summarization of open-ended survey responses \citep{wuttke2025ai,barari2025ai,landesvatter2026speech,mellon2024issues,halterman2025codebook,digiuseppe2026paired,han2025word}. & Often evaluates response quality, user experience, transcription, scaling, or coding agreement. The missing component is a randomized validation-item design for population and regression estimands. \\
Reference-sample and model-assisted inference & Uses fitted values for efficiency while reference observations protect the target estimand \citep{cochran1977sampling,sarndal1992model,wang2020methods,angelopoulos2023prediction,zrnic2024cross,egami2023imperfect,fan2024narratives,gronsbell2024anotherlook,chen2025unified,salerno2025modern,salerno2025moment,salerno2026spatial,visokay2025obesity}. & The usual setup has one fitted value and one reference value for each unit. Here validation is a person-variable or person-block matrix embedded in the interview, and variables may be outcomes or predictors. \\
\bottomrule
\end{tabularx}
\end{singlespace}
\end{table}

\section{Baseline Estimators}
\label{app:baseline-estimators}

In this section, we give the form of the baseline estimators for completeness. Using the notation from Section~\ref{sec:estimators}, an item-mean baseline uses only the validation-question answers and their known probabilities. The fixed-denominator version is the Horvitz-Thompson form of a two-phase survey estimator \citep{cochran1977sampling,sarndal1992model},
\[
  \thetahat^{HT}_j =
  \frac{1}{\Nhat_j}
  \sum_{i\in\mathcal I_n} w_iE_{ij}
  \frac{R_{ij}\Z_{ij}}{\pi_{ij}},
\]
where \(\Nhat_j=\sum_{i\in\mathcal I_n}w_iE_{ij}\). The corresponding H\'ajek estimate using only validation questions is the ratio,
\[
  \thetahat^{H}_j =
  \frac{\sum_{i\in\mathcal I_n} w_iE_{ij}R_{ij}\Z_{ij}/\pi_{ij}}
  {\sum_{i\in\mathcal I_n} w_iE_{ij}R_{ij}/\pi_{ij}}.
\]
These validation-question estimates are reported benchmarks. A fixed-denominator correction with no mapped-value term should be labeled separately from the ratio estimate based only on validation-question answers.

For a planned regression or other estimating-equation target, the baseline uses only records whose validation tile contains every structured response needed for the score. The score estimate using only those validation tiles solves
\[
  \sum_{i\in\mathcal I_n}w_iE_{iq}
  \frac{Q_{iq}}{\pi_{iq}}\psi_i(\beta)=0.
\]
The ratio-normalized version divides this score average by
\(\sum_{i\in\mathcal I_n}w_iE_{iq}Q_{iq}/\pi_{iq}\). The root is the same when the denominator does not depend on \(\beta\), but the distinction matters for how the score is scaled and how uncertainty is reported.

\section{Pairwise Moments for Linear Regressions}
\label{app:linear-moment-regressions}

For linear regressions, an alternative is to estimate the required moments. If \(\Z_j\) and \(\Z_k\) are both structured response variables, \(E_{ijk}=E_{ij}E_{ik}\), and \(\pi_{ijk}=\Prb(R_{ij}R_{ik}=1\mid \mathcal F_{ijk},E_{ijk}=1)\), a pairwise moment estimator using only validation-question pairs is
\[
  \widehat{\E}_w(\Z_j\Z_k)
  =
  \frac{1}{\Nhat_{jk}}
  \sum_{i\in\mathcal I_n} w_iE_{ijk}
  \frac{R_{ij}R_{ik}}{\pi_{ijk}}
  \Z_{ij}\Z_{ik},
  \qquad
  \Nhat_{jk}=\sum_{i\in\mathcal I_n}w_iE_{ijk}.
\]
A mapped-product version uses \(\Ztilde_{ij}\Ztilde_{ik}\) for everyone and corrects it with item pairs observed through validation questions:
\begin{equation}
  \widehat{\E}_w(\Z_j\Z_k)
  =
  \frac{1}{\Nhat_{jk}}
  \sum_{i\in\mathcal I_n}w_iE_{ijk}
  \left[
    \Ztilde_{ij}\Ztilde_{ik}
    +
    \frac{R_{ij}R_{ik}}{\pi_{ijk}}
    \{\Z_{ij}\Z_{ik}-\Ztilde_{ij}\Ztilde_{ik}\}
  \right].
  \label{eq:pairwise}
\end{equation}
Pairwise moments are useful for linear models when the needed variable pairs are observed together often enough. Nonlinear estimating equations usually require the relevant variables to be validated for the same respondent. Separately corrected pairwise moments can also produce a covariance matrix that needs projection or regularization before fitting planned linear regressions.
The ATUS regression examples use this linear-moment route with same-respondent planned validation blocks: the implementation corrects the \(X'X\) and \(X'Y\) moments, applies the fold-external calibration and tuning described in Section~\ref{sec:estimators}, and then solves the corrected normal equations.

\section{Additional Simulation and ATUS Results}
\label{app:additional-simulation-atus-results}

The following tables report numerical results summarized by the main-text simulation and ATUS figures. ATUS final weights are used as supplied in a pooled respondent-day estimand over the included survey years. The reference means and regression coefficients are therefore weighted pooled population-day summaries, with years contributing in proportion to their summed final weights. The complete diary is recoded into 32 reported structured response variables. Nineteen variables form the mutually exclusive 1,440-minute day composition, including sleep, non-sleep personal care, eating and drinking, paid work, commute travel, other travel, household activities, direct childcare, adult care, education, purchasing and services, social and leisure time, screen time, exercise, religious activity, volunteer and civic activity, telephone time, and other/uncodable minutes. Four additional minute variables describe time already counted in the 24-hour diary rather than adding new minutes: work at home, housework, food preparation, and secondary childcare. Nine participation indicators record whether the respondent had any paid work, work at home, commute, direct household childcare, direct non-household childcare, direct childcare, secondary childcare, adult care, or exercise.
Validation assignment is over \(p=250\) possible items: the 32 reported items plus 218 additional diary checks used only for assignment probabilities. A validation tile with \(B=18\) asks 7.2 percent of that validation universe. It represents an AMV design in which a conversational interview would be supplemented by a small set of structured validation items, rather than requiring every respondent to complete the full episode-level diary and associated follow-up questions.

\begin{table}[p]
\centering
\caption{Design-calibration simulation results at \(q=0.10\). Results are from 800 Monte Carlo repetitions with \(n=5{,}000\). Coverage uses nominal 95 percent linearization intervals. The simulation examines estimator behavior in a controlled setting and does not report an ATUS or health-surveillance empirical result.}
\label{tab:calibration}
\small
\begin{tabular}{lllrrr}
\toprule
Target & Estimator & Validation $q$ & Bias & RMSE & Coverage \\
\midrule
item mean & Mapping-only & 0.10 & -0.079 & 0.080 & 0.000 \\
item mean & Validation-question HT & 0.10 & -0.002 & 0.029 & 0.953 \\
item mean & Uncalibrated AMV item & 0.10 & -0.001 & 0.021 & 0.959 \\
item mean & Calibrated AMV item & 0.10 & -0.001 & 0.021 & 0.956 \\
regression slope & Mapping-only & 0.10 & 0.172 & 0.173 & 0.000 \\
regression slope & Validation-tile score & 0.10 & -0.001 & 0.082 & 0.969 \\
regression slope & Uncalibrated AMV score & 0.10 & -0.000 & 0.042 & 0.984 \\
regression slope & Calibrated AMV score & 0.10 & 0.000 & 0.042 & 0.976 \\
\bottomrule
\end{tabular}

\end{table}

\begin{table}[p]
\centering
\caption{ATUS item estimates under the moderate-error setting with 18 validation items from a 250-item validation universe. Reference is the weighted complete-diary reference estimand. Minute-variable quantities are in minutes, and direct-childcare participation quantities are in percentage points. Results summarize 30 validation-assignment repetitions under the simulated error mechanism. Effective validation \(n\) combines ATUS final weights with the realized validation-item design. Mean SE is the average linearized standard-error approximation used to compare estimators within the simulation; it should not be read as a calibrated coverage statement and does not include full ATUS replicate-weight sampling variance.}
\label{tab:atus-items}
\small
\begingroup
\setlength{\tabcolsep}{2pt}
\scriptsize
\begin{tabular}{l
  S[table-format=3.2]
  S[table-format=-3.2]
  S[table-format=-3.2]
  S[table-format=-3.2]
  S[table-format=-3.2]
  S[table-format=3.2]
  S[table-format=5.0]}
\toprule
 & {Reference} & {Mapping-only} & \multicolumn{2}{c}{Val. questions} & \multicolumn{2}{c}{AMV} & {Diagnostics} \\
\cmidrule(lr){2-2}\cmidrule(lr){3-3}\cmidrule(lr){4-5}\cmidrule(lr){6-7}\cmidrule(lr){8-8}
Structured response & {Mean} & {Bias} & {HT bias} & {Hájek bias} & {Calib. bias} & {Calib. RMSE} & {Eff. \(n\)} \\
\midrule
\multicolumn{8}{l}{\textit{Panel A: Minute variables}} \\
Sleep & 537.40 & -5.99 & 0.53 & 0.34 & 0.03 & 1.03 & 2113 \\
Paid work & 196.87 & 18.83 & -0.07 & -0.45 & 0.32 & 1.45 & 2108 \\
Commute & 14.80 & -3.49 & -0.03 & -0.04 & -0.10 & 0.43 & 2110 \\
Screen time & 197.26 & -12.12 & 0.38 & 0.11 & 0.10 & 1.50 & 2108 \\
Exercise & 19.85 & 3.45 & -0.15 & -0.06 & -0.11 & 0.45 & 2124 \\
Secondary childcare & 175.53 & 11.86 & -2.68 & -1.78 & -0.80 & 4.52 & 2118 \\
\addlinespace
\multicolumn{8}{l}{\textit{Panel B: Direct-childcare participation (percentage points)}} \\
Any direct childcare & 21.28 & -4.12 & 0.06 & 0.06 & -0.15 & 0.39 & 2108 \\
\bottomrule
\end{tabular}
\endgroup

\end{table}

\begin{table}[p]
\centering
\caption{ATUS regression preservation under the moderate-error setting with 18 validation items from a 250-item validation universe. Sleep effects are scaled to minutes of sleep per one-hour change in the structured predictor, and the direct-childcare coefficient is scaled to the percentage-point difference associated with children in the household. Outcome and marginal denote outcome-only correction and marginal plug-in correction. Validation-tile block is the block H\'ajek benchmark using only validation tiles. Uncal. AMV denotes uncalibrated AMV, the raw mapped moment correction without fold-external calibration or tuning. Calibrated AMV is the tuned fold-external calibrated moment estimator. Standard errors follow the convention in Appendix Table~\ref{tab:atus-items}.}
\label{tab:atus-regressions}
\small
\begingroup
\setlength{\tabcolsep}{1.8pt}
\scriptsize
\begin{tabular}{l
  S[table-format=-3.2]
  S[table-format=-3.2]
  S[table-format=-3.2]
  S[table-format=-3.2]
  S[table-format=-3.2]
  S[table-format=-3.2]
  S[table-format=-3.2]
  S[table-format=3.2]
  S[table-format=5.0]}
\toprule
 & {Reference} & \multicolumn{6}{c}{Bias by estimator} & \multicolumn{2}{c}{Diagnostics} \\
\cmidrule(lr){2-2}\cmidrule(lr){3-8}\cmidrule(lr){9-10}
Coefficient & {Effect} & {Mapping-only} & {Outcome} & {Marginal} & {Val. tile} & {Uncal. AMV} & {Calib. AMV} & {Calib. RMSE} & {Eff. \(n\)} \\
\midrule

\multicolumn{10}{l}{\textit{Panel A: Sleep regression (minutes of sleep per predictor hour)}} \\
Paid work & -12.08 & -4.26 & -0.37 & -2.69 & -0.09 & -0.07 & -0.12 & 0.66 & 1700 \\
Commute & -12.08 & -4.74 & -3.23 & 10.34 & 0.82 & 0.91 & 0.80 & 4.64 & 1700 \\
Screen time & -5.71 & -5.67 & -0.61 & -7.81 & 0.01 & 0.04 & -0.08 & 0.64 & 1700 \\
\addlinespace
\multicolumn{10}{l}{\textit{Panel B: Direct-childcare participation (percentage points)}} \\
Children in household & 50.82 & -14.19 & -0.41 & -0.45 & -0.76 & -0.42 & -0.52 & 1.23 & 1729 \\
\bottomrule
\end{tabular}
\endgroup

\end{table}

\begin{figure}[p]
\centering
\includegraphics[width=0.92\textwidth]{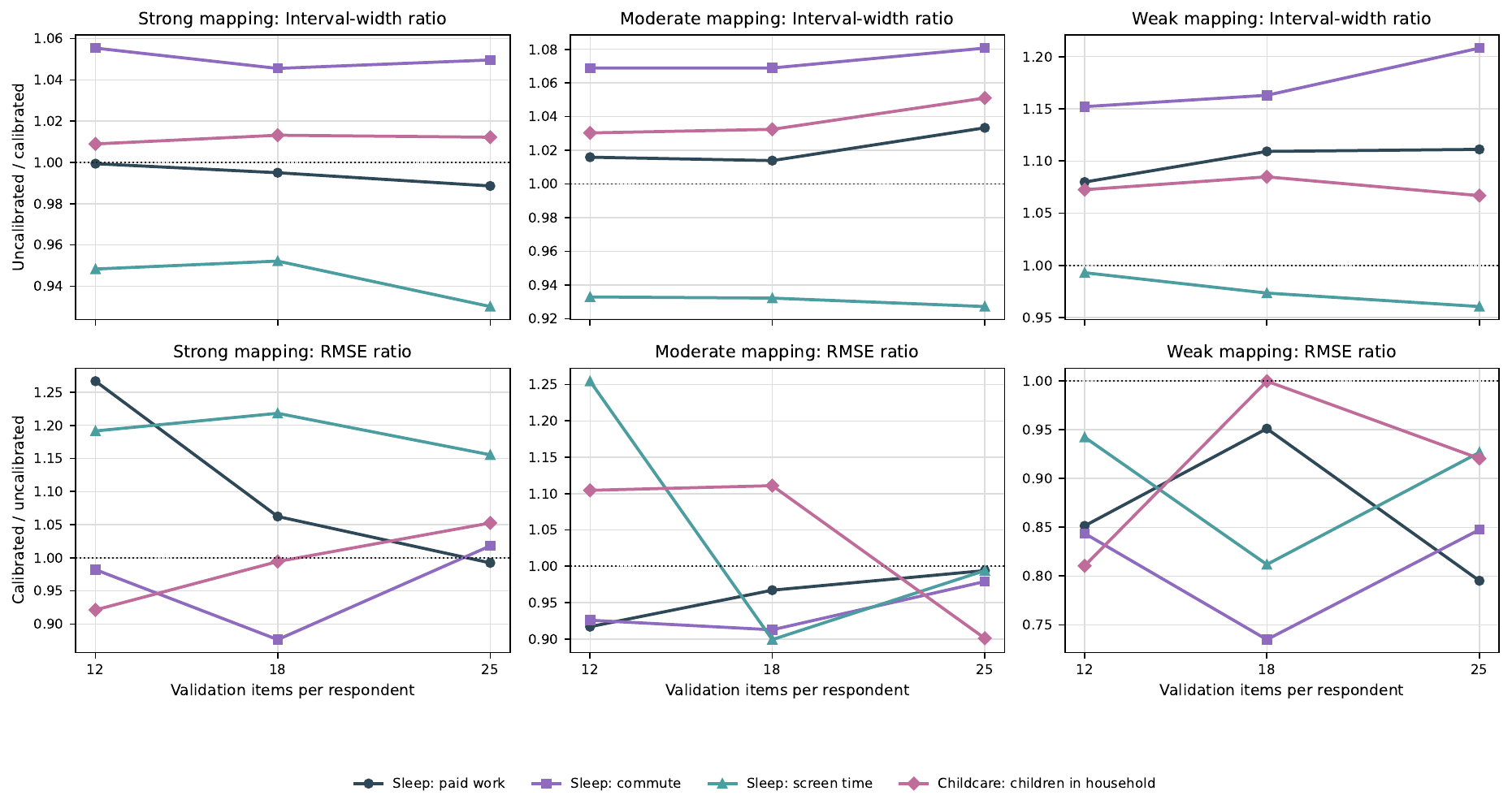}
\caption{Regression-only comparison of uncalibrated and calibrated AMV in the ATUS simulations. Uncalibrated AMV uses the raw mapped score without fold-external calibration or tuning. Calibrated AMV is the tuned fold-external moment estimator used in the main results. The top row plots the mean SE ratio, uncalibrated AMV divided by calibrated AMV. The bottom row plots the RMSE ratio, calibrated AMV divided by uncalibrated AMV. Values below one in the top row mean the uncalibrated diagnostic interval is narrower; values below one in the bottom row mean calibrated AMV has lower RMSE.}
\label{fig:atus-uncalibrated-comparison}
%\figalt{Six-panel line chart comparing uncalibrated and calibrated AMV for the planned ATUS regression coefficients across strong, moderate, and weak mapping settings. Uncalibrated intervals are often narrower, while calibrated AMV has lower RMSE for some coefficients and settings.}
\end{figure}

\begin{table}[p]
\centering
\caption{ATUS subgroup bias ranges under the moderate-error setting with 18 validation items from a 250-item validation universe. Ranges are over sex, employment, education, children-in-household, and weekday/weekend reporting groups. Sleep and secondary-childcare biases are in minutes; direct-childcare participation biases are in percentage points. Results summarize 30 validation-assignment repetitions under the simulated error mechanism.}
\label{tab:atus-subgroups}
\small

\begingroup
\setlength{\tabcolsep}{4pt}
\scriptsize
\begin{tabular}{l l l S[table-format=5.0]}
\toprule
Structured response & {Mapping-only bias range} & {Calib. AMV bias range} & {Min. eff. \(n\)} \\
\midrule
Sleep & -9.51 to -1.93 & -0.64 to 0.63 & 672 \\
Secondary childcare & -101.11 to 75.24 & -1.77 to -0.08 & 665 \\
Any direct childcare & -12.52 to 0.60 & -0.45 to 0.09 & 668 \\
\bottomrule
\end{tabular}
\endgroup

\end{table}

\begin{table}[p]
\centering
\caption{ATUS uncertainty and question-selection support. Panel A compares weighted absolute error in the bottom and top uncertainty deciles for the moderate scenario; direct-childcare participation is in percentage points and the other variables are in minutes. Panel B summarizes the main low-fraction design at \(B=18\) over \(p=250\) validation items. Panel C compares question-selection rules using scaled RMSE and mean scaled SE over the planned regression coefficients, using three validation-assignment repetitions for this auxiliary comparison. Effective \(n\) values combine ATUS final weights with realized propensities for the same-respondent validation sets.}
\label{tab:atus-design-support}
\small
\begingroup
\scriptsize
\setlength{\tabcolsep}{3pt}
\begin{tabularx}{\textwidth}{Xrrr}
\toprule
\multicolumn{4}{l}{Panel A: uncertainty calibration} \\
\midrule
Variable & Bottom decile & Top decile & Ratio \\
\midrule
Commute minutes & 4.63 & 14.60 & 3.16 \\
Secondary childcare minutes & 84.86 & 232.00 & 2.73 \\
Direct-childcare participation & 2.99 & 19.15 & 6.41 \\
\bottomrule
\end{tabularx}

\vspace{0.45em}

\begin{tabularx}{\textwidth}{Xrrrr}
\toprule
\multicolumn{5}{l}{Panel B: main low-fraction validation design} \\
\midrule
Quantity & Cells/items & Fraction & Eff. \(n\) & Cal./direct SE \\
\midrule
Validation universe & 250 & -- & -- & -- \\
Validation burden per respondent & 18 & 0.072 & -- & -- \\
Priority item checks, range & 7 & 0.074--0.074 & 2,108--2,124 & 0.36--0.65 \\
Sleep regression block & 4 & 0.060 & 1,700 & 0.66--0.91 \\
Childcare regression block & 2 & 0.061 & 1,729 & 0.63--0.71 \\
\bottomrule
\end{tabularx}

\vspace{0.45em}

\begin{tabularx}{\textwidth}{Xrrrr}
\toprule
\multicolumn{5}{l}{Panel C: question-selection comparisons} \\
\midrule
Question-selection rule & Item RMSE & Cal. reg. RMSE & Direct reg. RMSE & Cal./direct SE \\
\midrule
Full low-fraction rule & 0.0192 & 1.084 & 2.175 & 0.81 \\
Planned regression block 8\% & 0.0142 & 1.112 & 2.210 & 0.80 \\
Planned regression block 10\% & 0.0159 & 1.762 & 2.586 & 0.80 \\
Random only & 0.0195 & 96.180 & 96.180 & 1.00 \\
No planned regression block & 0.0301 & 0.438 & 0.438 & 1.00 \\
\bottomrule
\end{tabularx}
\endgroup

\end{table}

\section{Local-Model ATUS Interview-Coding Check}
\label{app:atus-llm-validation}

This appendix gives empirical content to the strong, moderate, and weak error regimes used in the ATUS simulation. We selected 100 ATUS respondent-days from the same 2018, 2019, and 2021--2024 source years used in Section~\ref{sec:atus}. Each complete diary was rewritten as hidden respondent context containing episode order, activity durations, locations, and childcare flags. A local interview-coding exercise then generated fixed interview questions from the diary facts, generated answers in natural language, and coded the transcript into the 32 ATUS structured variables, a confidence score, and an inferability label for each variable.

The comparison target is the complete ATUS diary reconstruction used elsewhere in the paper. For variable \(j\), the appendix reports bias \(\bar{\tilde Z}_j-\bar Z_j\), MAE, RMSE, and the unexplained-variation ratio \(\sum_i(\tilde Z_{ij}-Z_{ij})^2/n\) divided by the sample variance of \(Z_j\). This is the same scale used in Section~\ref{sec:sample-size}; one minus this ratio is the share of item-level variation explained by the mapped value, relative to using only the item mean. The unexplained-variation quartiles define empirical anchors for the strong, moderate, and weak regimes. This protocol is deliberately modest: one fixed interview, no software search, and no human correction of transcript coding. The high unexplained-variation ratios are an empirical finding from this protocol. They place this simple local setup near the weak side of the design problem and explain why the main paper reports a controlled range of accuracy rather than relying on a single transcript-generation run.

\begin{table}[!htbp]
\centering
\caption{Local-model ATUS interview mapping check for priority variables}
\label{tab:atus_llm_validation_priority}
\begin{tabular}{lrrrrrr}
\toprule
Variable & Truth mean & Coded mean & Bias & MAE & RMSE & Residual ratio \\
\midrule
Any direct childcare & 0.380 & 0.480 & 0.100 & 0.175 & 0.409 & 0.71 \\
Commute & 15.2 & 47.4 & 32.2 & 44.0 & 66.1 & 3.71 \\
Exercise & 25.3 & 27.4 & 2.1 & 14.5 & 34.7 & 0.33 \\
Paid work & 217.7 & 133.5 & -84.2 & 92.5 & 171.1 & 0.46 \\
Screen time & 172.2 & 55.4 & -116.8 & 131.4 & 187.0 & 1.22 \\
Secondary childcare & 272.2 & 7.3 & -264.9 & 269.5 & 534.4 & 1.33 \\
Sleep & 518.7 & 461.8 & -56.9 & 191.8 & 271.1 & 4.44 \\
\bottomrule
\end{tabular}
\begin{minipage}{0.92\linewidth}
\footnotesize
\emph{Notes:} The study uses 100 ATUS respondent-days from 2018, 2019, and 2021--2024. gemma4:latest produced the fixed interview questions and mistral:latest answered from hidden ATUS diary context. mistral:latest coded the transcript into the 32 ATUS formal variables. Residual ratio is mean squared coding error divided by the sample variance of the complete-diary value. Minute rows are minutes per respondent-day; binary rows use probabilities.
\end{minipage}
\end{table}

\begin{table}[!htbp]
\centering
\caption{Empirical anchors for strong, moderate, and weak mapping regimes}
\label{tab:atus_llm_hml_anchors}
\begin{tabular}{llrr}
\toprule
Regime & Empirical definition & All variables & Priority variables \\
\midrule
Strong & best quartile within the local-model run & 0.64 & 0.58 \\
Moderate & median within the local-model run & 1.11 & 1.22 \\
Weak & upper quartile within the local-model run & 1.81 & 2.52 \\
\bottomrule
\end{tabular}
\begin{minipage}{0.92\linewidth}
\footnotesize
\emph{Notes:} Anchors are quantiles of variable-level residual ratios from the local-model interview check. These values are used to interpret the strong, moderate, and weak settings in the main ATUS simulation.
\end{minipage}
\end{table}

\begin{figure}[p]
\centering
\includegraphics[width=0.86\textwidth]{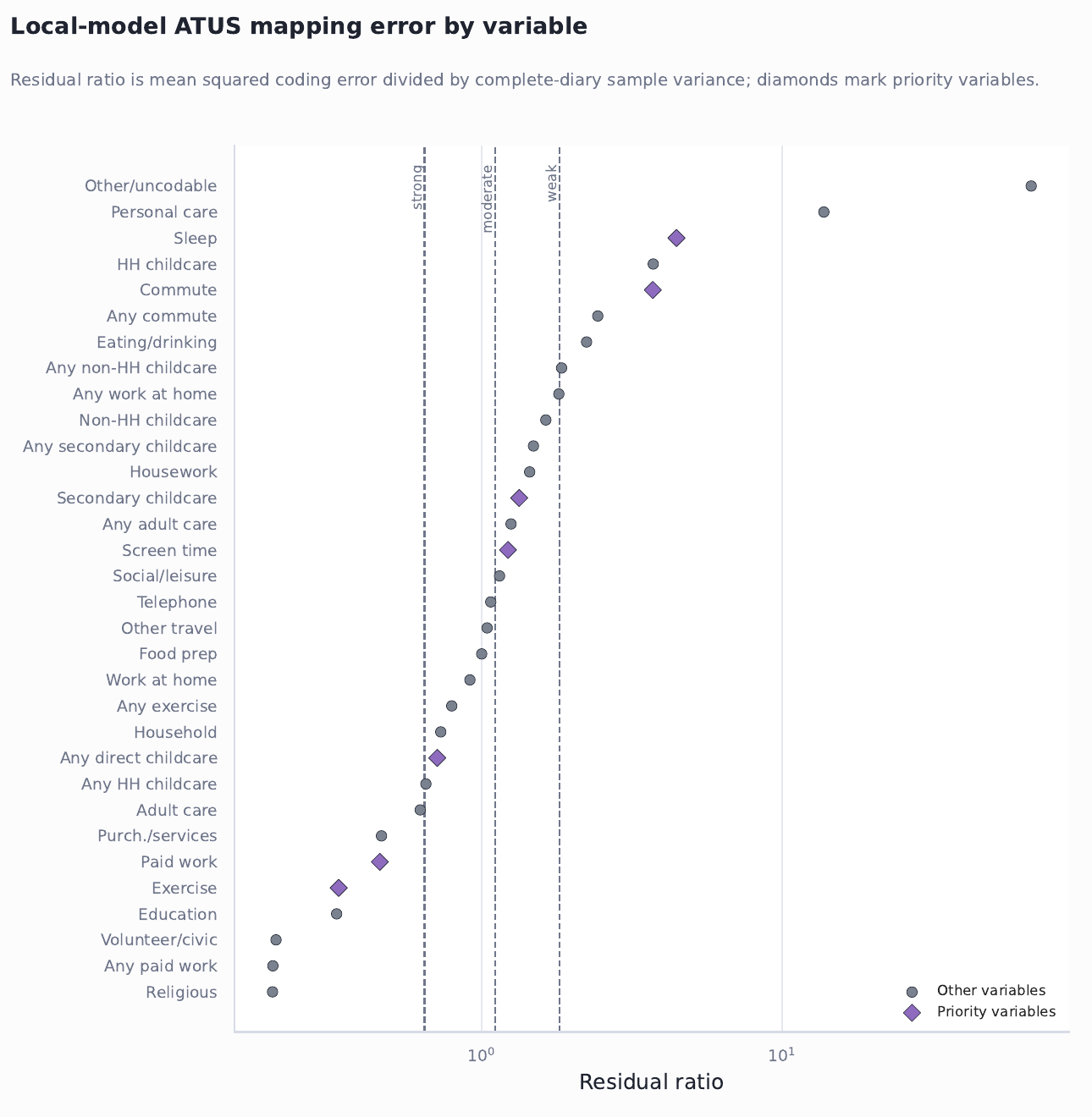}
\caption{Variable-level unexplained variation from the local-model ATUS interview-coding check. The unexplained-variation ratio is mean squared coding error divided by the sample variance of the complete-diary value; one minus this ratio is the share of item-level variation explained by the mapped value, relative to using only the item mean. Diamond points mark the seven priority variables used in the main ATUS RMSE display. Values above one indicate that transcript coding leaves more squared error than a marginal-mean baseline for that variable.}
\label{fig:atus-llm-residual-ratios}
%\figalt{Variable-level unexplained-variation plot for the 32 ATUS structured variables. Several high-variance duration variables, including sleep and commute, have unexplained-variation ratios above one, while exercise and paid work are below one.}
\end{figure}

\begin{figure}[p]
\centering
\includegraphics[width=\textwidth]{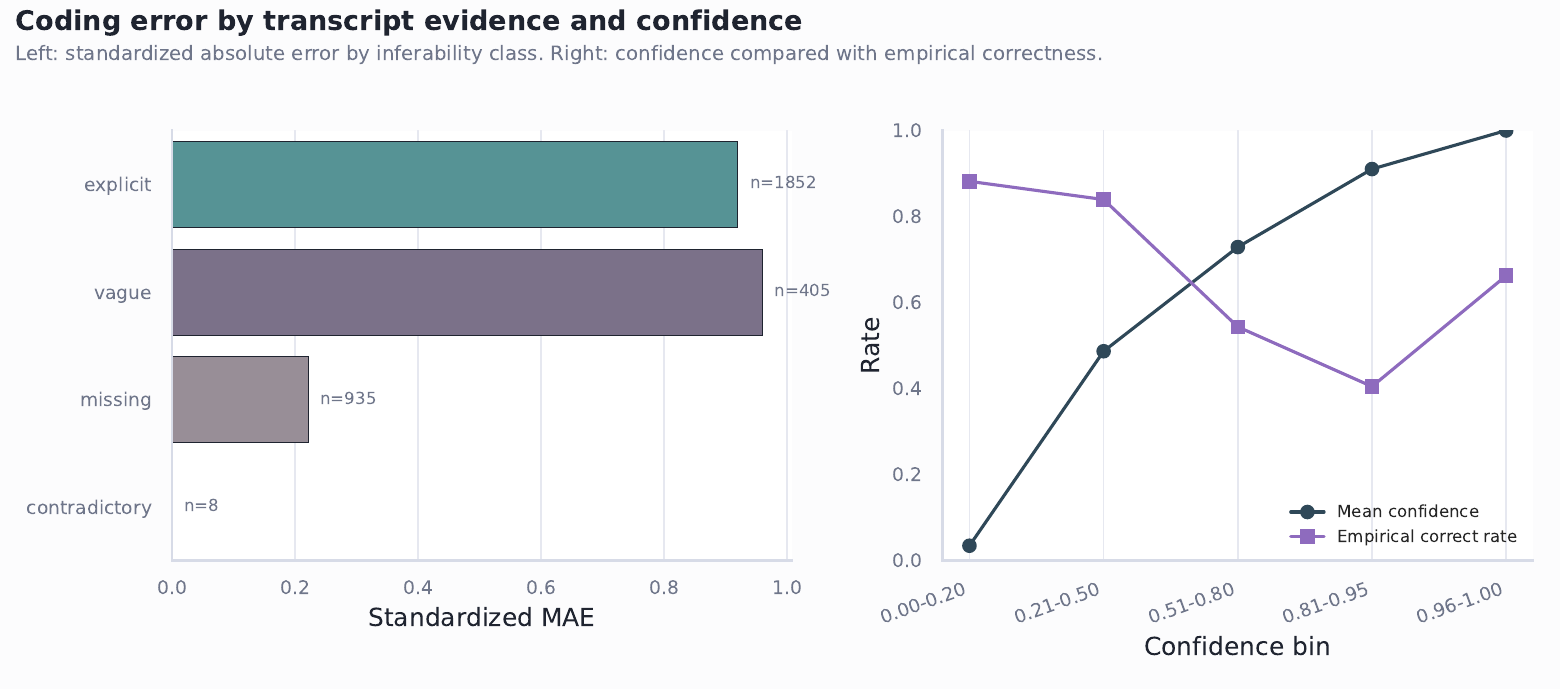}
\caption{Inferability and confidence checks from the local-model ATUS interview-coding check. The left panel summarizes standardized MAE by the coding model's inferability label. The right panel compares mean confidence with the empirical fraction of values coded within a variable-specific tolerance. The confidence scores are descriptive outputs from the coding model and are not used as calibrated probabilities in the main ATUS simulation.}
\label{fig:atus-llm-inferability-calibration}
%\figalt{Two-panel figure. Explicit and vague inferability labels have larger standardized errors than missing labels, and high confidence bins do not line up monotonically with the empirical fraction of items coded within tolerance.}
\end{figure}
\FloatBarrier

\section{Additional Details on CHAMPS Data}
\label{app:champs-data}

This appendix gives reproducibility details for the CHAMPS analysis using existing narratives and structured verbal-autopsy responses. The analysis used the restricted de-identified CHAMPS L2 export dated 2026-06-01. Source records were treated as read-only. The summaries below were computed from the de-identified verbal-autopsy file, and the derived outputs store only aggregate narrative-length summaries and figures, not narrative text.

The verbal-autopsy file contains 9,299 case records. The designated de-identified narrative field is nonempty for 4,693 cases, or 50.5 percent of records. Among nonempty narratives, the median length is 108 whitespace-delimited words and the mean length is 132 words. The interquartile range is 67 to 165 words, with a long right tail reaching 867 words. These lengths indicate that actual narrative text is available for a substantial subset of cases. The analysis reports narrative availability and treats the blank-narrative subset separately rather than assuming universal text coverage \citep{champsdata2025}.

\begin{figure}[p]
\centering
\includegraphics[width=\textwidth]{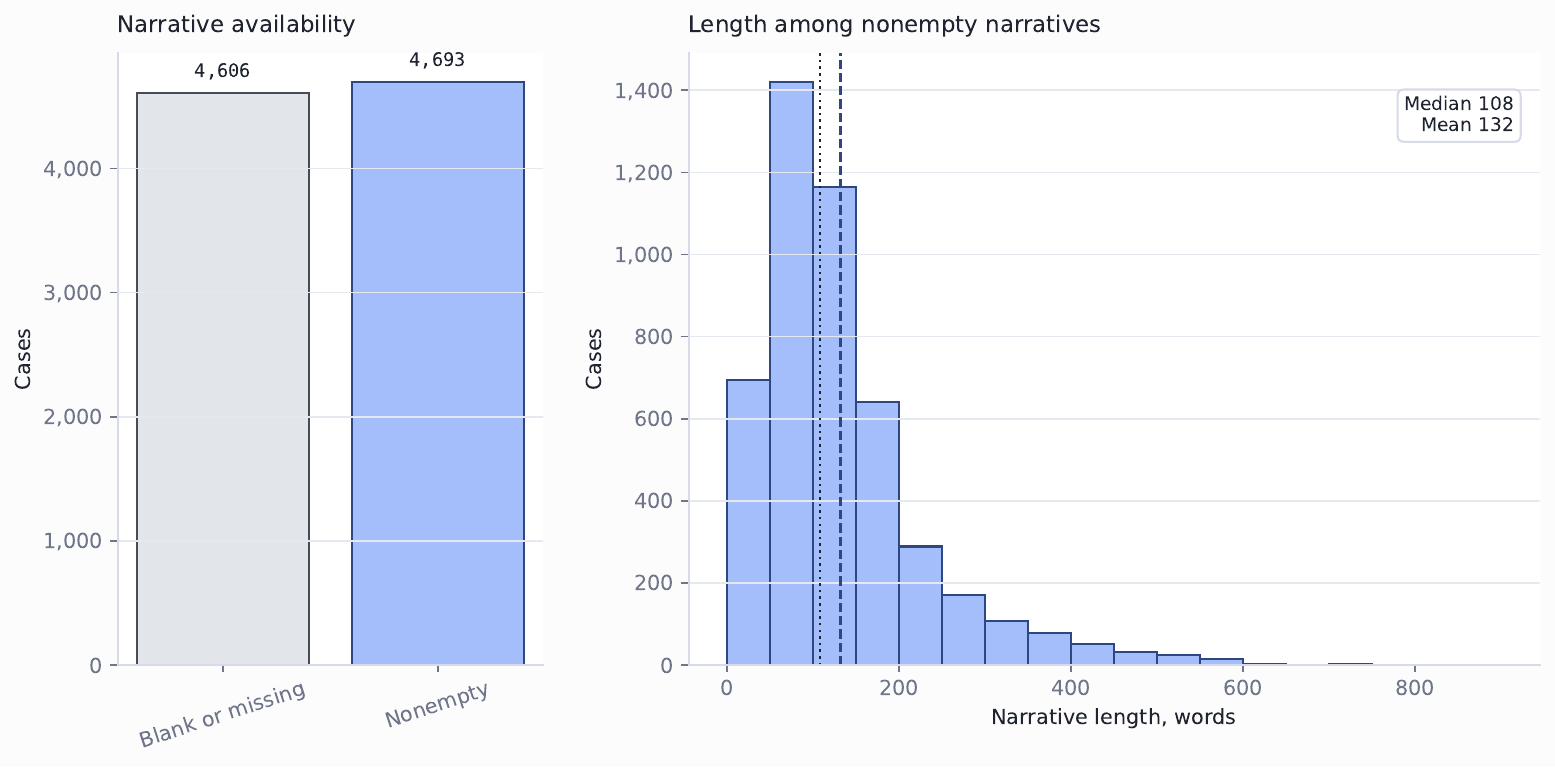}
\caption{CHAMPS verbal-autopsy narrative availability and word length. The left panel shows blank or missing versus nonempty narrative fields across all 9,299 verbal-autopsy case records. The right panel shows the word-count distribution among the 4,693 cases with nonempty narratives. No narrative text is stored in the derived figure or summary table.}
\label{fig:champs-narrative-length}
%\figalt{Two-panel figure. The left panel shows nearly equal counts of blank or missing and nonempty narrative fields, with 4,606 blank or missing and 4,693 nonempty. The right panel is a right-skewed histogram of nonempty narrative word counts, with most narratives below 200 words and a long tail approaching 900 words.}
\end{figure}

Figure~\ref{fig:champs-narrative-length} displays the same availability split and the nonempty narrative length distribution.

\begin{table}[p]
\centering
\caption{CHAMPS verbal-autopsy narrative availability and length. Length is counted in whitespace-delimited words from the de-identified verbal-autopsy narrative field in the restricted CHAMPS L2 export dated 2026-06-01. Summary statistics for word length are conditional on the narrative field being nonempty.}
\label{tab:champs-narrative-summary}
\small
\begin{tabular}{lr}
\toprule
Quantity & Value \\
\midrule
Cases in verbal-autopsy file & 9,299 \\
Cases with nonempty narrative & 4,693 \\
Cases with blank or missing narrative field & 4,606 \\
Percent with nonempty narrative & 50.5\% \\
Mean words among nonempty narratives & 132 \\
Median words among nonempty narratives & 108 \\
Interquartile range, words & 67--165 \\
95th percentile, words & 333 \\
Maximum words & 867 \\
\bottomrule
\end{tabular}

\end{table}

\subsection{Selected CHAMPS example estimates}

Table~\ref{tab:champs-items-app} reports the numeric structured-response fraction results shown in Figure~\ref{fig:champs-items} for nineteen selected constructs. Appendix Table~\ref{tab:champs-item-support-grid} reports the item-calibration support rule used for the displayed results and two nearby alternatives. Appendix Table~\ref{tab:champs-proxy-diagnostics} reports aggregate diagnostics for the fixed Narrative comparator. Appendix Table~\ref{tab:champs-goldilocks-mentions} separately reports direct mention rates for six indicators that are often omitted from short narratives.

\begin{table}[p]
\centering
\caption{Appendix numeric values for selected CHAMPS structured-response fractions. Results use records with nonempty narratives and observed structured responses for each construct. Structured VA is the structured-verbal-autopsy fraction among those records. Narrative uses fixed phrase and duration rules applied to the narrative text. Validation only is the H\'ajek estimate using revealed structured-verbal-autopsy answers. Uncal. AMV uses the raw narrative-derived mapped value in \eqref{eq:item-cal-amv}, with no fold-tuned shrinkage. Cal. AMV uses the item estimator in \eqref{eq:item-cal-amv}, with fold-external \(\lambda\) calibration and validation-weighted correction from revealed structured-verbal-autopsy answers. All validation estimates use the sparse validation rule over 100 validation-assignment draws.}
\label{tab:champs-items-app}
\footnotesize
\resizebox{\textwidth}{!}{%
\begin{tabular}{lrrrrrrrrrrrrrrrrrr}
\toprule
Construct & Structured VA & Narrative & Validation only & Val. error & Val. SE & Val. RMSE & Uncal. AMV & Uncal. error & Uncal. SE & Uncal. RMSE & Cal. AMV & Cal. error & Cal. SE & Cal. RMSE & $\lambda$ & Supported & Revealed $n$ & Revealed frac. \\
\midrule
Motorized transport to facility & 0.894 & 0.183 & 0.891 & -0.002 & 0.022 & 0.020 & 0.894 & 0.000 & 0.058 & 0.058 & 0.892 & -0.002 & 0.022 & 0.020 & 0.017 & 0.95 & 199 & 0.090 \\
Smaller than usual at birth & 0.813 & 0.100 & 0.810 & -0.003 & 0.039 & 0.038 & 0.812 & -0.001 & 0.081 & 0.079 & 0.810 & -0.003 & 0.039 & 0.039 & 0.029 & 0.91 & 101 & 0.090 \\
Treatment received during illness & 0.697 & 0.487 & 0.693 & -0.004 & 0.029 & 0.026 & 0.697 & 0.000 & 0.038 & 0.039 & 0.693 & -0.003 & 0.029 & 0.027 & 0.227 & 1.00 & 243 & 0.090 \\
Respiratory symptoms & 0.692 & 0.429 & 0.689 & -0.003 & 0.029 & 0.027 & 0.690 & -0.002 & 0.037 & 0.035 & 0.689 & -0.003 & 0.028 & 0.026 & 0.268 & 1.00 & 255 & 0.090 \\
Traveled to health facility & 0.629 & 0.557 & 0.631 & 0.002 & 0.027 & 0.025 & 0.631 & 0.002 & 0.036 & 0.037 & 0.631 & 0.002 & 0.027 & 0.025 & 0.111 & 1.00 & 318 & 0.090 \\
Care sought outside home & 0.480 & 0.902 & 0.479 & -0.001 & 0.031 & 0.033 & 0.478 & -0.002 & 0.043 & 0.041 & 0.479 & -0.001 & 0.031 & 0.032 & 0.049 & 1.00 & 252 & 0.090 \\
Illness duration at least 3 days & 0.416 & 0.533 & 0.415 & -0.001 & 0.031 & 0.027 & 0.418 & 0.002 & 0.040 & 0.039 & 0.416 & -0.001 & 0.031 & 0.026 & 0.124 & 1.00 & 262 & 0.091 \\
Fever/infectious symptoms & 0.393 & 0.353 & 0.392 & -0.001 & 0.031 & 0.030 & 0.393 & -0.001 & 0.026 & 0.025 & 0.393 & -0.001 & 0.024 & 0.022 & 0.666 & 1.00 & 237 & 0.089 \\
Gastrointestinal/diarrheal symptoms & 0.322 & 0.312 & 0.325 & 0.002 & 0.029 & 0.029 & 0.323 & 0.001 & 0.032 & 0.028 & 0.324 & 0.002 & 0.027 & 0.026 & 0.411 & 1.00 & 251 & 0.090 \\
Neurologic symptoms & 0.282 & 0.145 & 0.283 & 0.001 & 0.028 & 0.027 & 0.281 & -0.002 & 0.029 & 0.028 & 0.282 & -0.001 & 0.027 & 0.024 & 0.448 & 1.00 & 253 & 0.091 \\
Care costs displaced household payments & 0.242 & 0.045 & 0.242 & -0.000 & 0.024 & 0.024 & 0.242 & -0.001 & 0.029 & 0.028 & 0.242 & -0.001 & 0.024 & 0.024 & 0.014 & 1.00 & 316 & 0.090 \\
Vomiting & 0.240 & 0.207 & 0.236 & -0.005 & 0.027 & 0.027 & 0.242 & 0.002 & 0.025 & 0.023 & 0.240 & -0.001 & 0.023 & 0.022 & 0.582 & 1.00 & 251 & 0.091 \\
Illness duration at least 7 days & 0.217 & 0.455 & 0.217 & 0.000 & 0.026 & 0.024 & 0.216 & -0.000 & 0.039 & 0.044 & 0.217 & 0.000 & 0.026 & 0.024 & 0.084 & 1.00 & 260 & 0.090 \\
Convulsions or fits & 0.160 & 0.102 & 0.159 & -0.001 & 0.023 & 0.021 & 0.160 & 0.000 & 0.023 & 0.022 & 0.160 & -0.000 & 0.021 & 0.018 & 0.510 & 1.00 & 248 & 0.090 \\
Cough & 0.151 & 0.087 & 0.155 & 0.004 & 0.023 & 0.022 & 0.154 & 0.003 & 0.021 & 0.019 & 0.154 & 0.003 & 0.020 & 0.019 & 0.612 & 1.00 & 247 & 0.090 \\
Problems with treatment or respect & 0.150 & 0.009 & 0.147 & -0.003 & 0.025 & 0.022 & 0.147 & -0.003 & 0.027 & 0.027 & 0.147 & -0.003 & 0.025 & 0.023 & 0.092 & 1.00 & 199 & 0.090 \\
More than 2 hours to nearest facility & 0.148 & 0.019 & 0.147 & -0.001 & 0.020 & 0.019 & 0.148 & 0.000 & 0.022 & 0.023 & 0.147 & -0.001 & 0.020 & 0.019 & 0.038 & 1.00 & 319 & 0.091 \\
Doubts medical care was needed & 0.084 & 0.003 & 0.083 & -0.001 & 0.015 & 0.015 & 0.083 & -0.002 & 0.016 & 0.017 & 0.083 & -0.001 & 0.016 & 0.015 & 0.032 & 0.98 & 315 & 0.090 \\
Traditional medicine used & 0.080 & 0.029 & 0.081 & 0.000 & 0.015 & 0.014 & 0.080 & -0.001 & 0.014 & 0.014 & 0.080 & -0.000 & 0.014 & 0.013 & 0.605 & 0.99 & 316 & 0.090 \\
\bottomrule
\end{tabular}%
}

\end{table}

\begin{table}[p]
\centering
\caption{CHAMPS item-calibration support rules for the displayed item estimates. The chosen rule requires effective validation \(n\ge 30\), effective positive \(n\ge 10\), and effective negative \(n\ge 10\) overall and in each training fold. SE and RMSE ratios compare calibrated AMV with validation questions only across the nineteen displayed constructs; values below one favor calibrated AMV. The support column records the share of validation-assignment draws in which the fold-specific rule was met; unsupported draws fall back to validation only.}
\label{tab:champs-item-support-grid}
\footnotesize
\resizebox{\textwidth}{!}{%
\begin{tabular}{lrrrrrrrrrrr}
\toprule
Rule & Min $n_{eff}$ & Min pos. & Min neg. & Support & $\lambda$ & Mean SE ratio & Median SE ratio & SE wins & Mean RMSE ratio & Median RMSE ratio & RMSE wins \\
\midrule
Default (chosen) & 30 & 10 & 10 & 0.99 & 0.259 & 0.957 & 0.998 & 11 & 0.950 & 0.990 & 11 \\
Lenient & 20 & 8 & 8 & 1.00 & 0.259 & 0.957 & 0.998 & 11 & 0.951 & 0.990 & 11 \\
Strict & 50 & 15 & 15 & 0.91 & 0.260 & 0.957 & 0.998 & 11 & 0.951 & 0.990 & 11 \\
\bottomrule
\end{tabular}%
}

\end{table}

\begin{table}[p]
\centering
\caption{Aggregate diagnostics for the CHAMPS Narrative comparator. The comparator uses fixed phrase and duration rules applied to the narrative text. It does not use structured-response targets or case variables. Errors are differences between Narrative and Structured VA fractions across the displayed constructs.}
\label{tab:champs-proxy-diagnostics}
\small
\begin{tabular}{lr}
\toprule
Diagnostic & Value \\
\midrule
proxy\_inputs & fixed\_phrase\_and\_duration\_rules \\
case\_variables\_used & false \\
structured\_targets\_used\_to\_build\_proxy & false \\
displayed\_constructs & 19.000 \\
mean\_abs\_proxy\_error & 0.194 \\
max\_abs\_proxy\_error & 0.713 \\
constructs\_abs\_error\_ge\_0\_10 & 11.000 \\
constructs\_abs\_error\_ge\_0\_20 & 6.000 \\
\bottomrule
\end{tabular}

\end{table}

Table~\ref{tab:champs-regression-app} reports the numeric regression results shown in Figure~\ref{fig:champs-regression}. Table~\ref{tab:champs-regression-search} reports the displayed models and comparison models considered while selecting the two examples.

\begin{table}[p]
\centering
\caption{Appendix numeric values for selected CHAMPS regression estimates. The traditional-medicine model uses observed case variables as predictors. The treatment-received model uses five mapped predictors and includes age group and site controls, which are not displayed in the main-text panel. Narrative uses fixed phrase and duration rules applied to the narrative text, without validation items. Validation only uses the revealed same-respondent validation questions. Uncal. AMV uses the raw mapped regression score with the validation-weighted correction, but no fold-tuned score shrinkage. Cal. AMV uses the calibrated score equation in \eqref{eq:score-cal-amv}, with scalar fold-external score calibration under the sparse validation rule. Spread is the across-assignment standard deviation from the 400 validation-assignment draws.}
\label{tab:champs-regression-app}
\footnotesize
\resizebox{\textwidth}{!}{%
\begin{tabular}{llrrrrrrrrrrrrrrrr}
\toprule
Model & Coefficient & Structured VA & Structured SE & Narrative & Narrative SE & Validation only & Val. spread & Val. RMSE & Uncal. AMV & Uncal. spread & Uncal. RMSE & Cal. AMV & Cal. spread & Cal. RMSE & Cal. error & Cal./Val. spread & Cal./Uncal. spread \\
\midrule
Traditional medicine used & Site: Kenya & 0.092 & 0.012 & 0.021 & 0.006 & 0.090 & 0.035 & 0.035 & 0.092 & 0.034 & 0.034 & 0.091 & 0.033 & 0.033 & -0.001 & 0.95 & 0.96 \\
Traditional medicine used & Site: Sierra Leone & 0.032 & 0.013 & 0.021 & 0.008 & 0.030 & 0.039 & 0.039 & 0.031 & 0.036 & 0.036 & 0.031 & 0.035 & 0.035 & -0.001 & 0.91 & 0.98 \\
Traditional medicine used & Sex: not male & 0.009 & 0.009 & -0.006 & 0.006 & 0.008 & 0.025 & 0.025 & 0.008 & 0.024 & 0.024 & 0.008 & 0.023 & 0.023 & -0.001 & 0.92 & 0.96 \\
Traditional medicine used & Death outside facility & -0.002 & 0.015 & 0.025 & 0.011 & -0.006 & 0.042 & 0.042 & -0.002 & 0.035 & 0.035 & -0.004 & 0.035 & 0.035 & -0.002 & 0.83 & 0.99 \\
Traditional medicine used & Age: infant & -0.025 & 0.020 & -0.004 & 0.014 & -0.018 & 0.055 & 0.055 & -0.021 & 0.050 & 0.050 & -0.020 & 0.049 & 0.049 & 0.005 & 0.88 & 0.97 \\
Traditional medicine used & Site: South Africa & -0.033 & 0.011 & -0.014 & 0.006 & -0.035 & 0.030 & 0.030 & -0.033 & 0.028 & 0.028 & -0.034 & 0.027 & 0.027 & -0.001 & 0.91 & 0.98 \\
Traditional medicine used & Site: Nigeria & -0.040 & 0.027 & -0.010 & 0.013 & -0.016 & 0.038 & 0.045 & -0.040 & 0.027 & 0.027 & -0.039 & 0.033 & 0.033 & 0.001 & 0.86 & 1.22 \\
Traditional medicine used & Stillbirth & -0.104 & 0.019 & -0.056 & 0.011 & -0.101 & 0.050 & 0.050 & -0.101 & 0.051 & 0.051 & -0.100 & 0.048 & 0.048 & 0.003 & 0.96 & 0.94 \\
Traditional medicine used & Age: late neonate & -0.111 & 0.018 & -0.041 & 0.013 & -0.105 & 0.051 & 0.051 & -0.107 & 0.050 & 0.050 & -0.106 & 0.047 & 0.047 & 0.006 & 0.92 & 0.94 \\
Traditional medicine used & Age: first 24 hours & -0.117 & 0.018 & -0.056 & 0.011 & -0.114 & 0.049 & 0.049 & -0.115 & 0.049 & 0.049 & -0.114 & 0.046 & 0.046 & 0.003 & 0.95 & 0.95 \\
Traditional medicine used & Age: early neonate & -0.125 & 0.016 & -0.051 & 0.011 & -0.121 & 0.044 & 0.044 & -0.124 & 0.043 & 0.043 & -0.122 & 0.040 & 0.040 & 0.003 & 0.92 & 0.94 \\
Treatment received & Difficulty breathing & 0.188 & 0.018 & 0.027 & 0.026 & 0.189 & 0.040 & 0.040 & 0.187 & 0.076 & 0.076 & 0.190 & 0.041 & 0.041 & 0.002 & 1.02 & 0.54 \\
Treatment received & Fever/infectious symptoms & 0.138 & 0.024 & 0.149 & 0.024 & 0.136 & 0.053 & 0.053 & 0.134 & 0.083 & 0.083 & 0.136 & 0.052 & 0.052 & -0.002 & 0.99 & 0.63 \\
Treatment received & Vomiting & 0.070 & 0.021 & 0.121 & 0.023 & 0.071 & 0.048 & 0.048 & 0.077 & 0.071 & 0.072 & 0.071 & 0.047 & 0.047 & 0.002 & 0.99 & 0.66 \\
Treatment received & Traditional medicine & 0.067 & 0.027 & 0.080 & 0.046 & 0.068 & 0.063 & 0.063 & 0.066 & 0.077 & 0.077 & 0.067 & 0.062 & 0.062 & 0.001 & 0.99 & 0.81 \\
Treatment received & Cough & 0.034 & 0.022 & 0.151 & 0.030 & 0.036 & 0.048 & 0.048 & 0.037 & 0.071 & 0.071 & 0.036 & 0.048 & 0.048 & 0.002 & 1.00 & 0.69 \\
\bottomrule
\end{tabular}%
}

\end{table}

Table~\ref{tab:champs-regression-support} reports the realized validation support for the two displayed regressions.

\begin{table}[p]
\centering
\caption{Realized validation support for the displayed CHAMPS regressions. Revealed \(n\) is the average number of records with the validation questions needed for the working regression. Share all is the average share of all nonempty-narrative records with those validation questions revealed. Share eligible is the corresponding share among records with all structured responses needed for the regression. Supported is the share of validation-assignment draws with enough records to fit calibrated score AMV. The final columns compare calibrated AMV with validation only and uncalibrated AMV; ratios below one favor calibrated AMV.}
\label{tab:champs-regression-support}
\small
\resizebox{\textwidth}{!}{%
\begin{tabular}{llrrrrrrrrrrr}
\toprule
Model & Outcome & Formal pred. & Controls & Complete $n$ & Revealed $n$ & Share all & Share eligible & Supported & Cal./Val. spread & Cal./Uncal. spread & Cal./Val. RMSE & Cal./Uncal. RMSE \\
\midrule
Traditional medicine used & traditional\_medicine & 0 & 4 & 3,522 & 424 & 0.090 & 0.121 & 1.00 & 0.91 & 0.98 & 0.90 & 0.99 \\
Treatment received & treatment\_received & 5 & 2 & 2,459 & 421 & 0.090 & 0.171 & 1.00 & 1.00 & 0.67 & 1.00 & 0.67 \\
\bottomrule
\end{tabular}%
}

\end{table}

\begin{table}[p]
\centering
\caption{CHAMPS regressions considered for the displayed examples. Formal pred. is the number of structured predictors translated from the narrative. Controls is the number of observed case-variable blocks. Complete \(n\) is the number of nonempty-narrative records with the outcome and all formal predictors observed. Supported is the share of validation-assignment draws in which calibrated score AMV was supported. The final columns compare calibrated AMV with validation only and uncalibrated AMV; ratios above one indicate no precision gain for that comparison.}
\label{tab:champs-regression-search}
\footnotesize
\resizebox{\textwidth}{!}{%
\begin{tabular}{llrrrrrrrrrrr}
\toprule
Model & Outcome & Formal pred. & Controls & Complete $n$ & Revealed $n$ & Share all & Share eligible & Supported & Cal./Val. spread & Cal./Uncal. spread & Cal./Val. RMSE & Cal./Uncal. RMSE \\
\midrule
Traditional medicine used & traditional\_medicine & 0 & 4 & 3,522 & 424 & 0.090 & 0.121 & 1.00 & 0.91 & 0.98 & 0.90 & 0.99 \\
Treatment received & treatment\_received & 5 & 2 & 2,459 & 421 & 0.090 & 0.171 & 1.00 & 1.00 & 0.67 & 1.00 & 0.67 \\
Convulsions or fits & convulsions & 0 & 4 & 2,741 & 423 & 0.090 & 0.154 & 1.00 & 0.96 & 0.84 & 0.95 & 0.84 \\
Vomiting & vomiting & 0 & 4 & 2,775 & 423 & 0.090 & 0.152 & 1.00 & 0.96 & 0.81 & 0.96 & 0.81 \\
Care sought outside home & care\_sought\_outside\_home & 5 & 0 & 2,464 & 425 & 0.090 & 0.172 & 1.00 & 1.00 & 0.63 & 1.00 & 0.62 \\
Treatment received & treatment\_received & 4 & 2 & 2,478 & 423 & 0.090 & 0.171 & 1.00 & 1.01 & 0.63 & 1.01 & 0.63 \\
\bottomrule
\end{tabular}%
}

\end{table}

Table~\ref{tab:champs-goldilocks-mentions} gives a simple aggregate check on whether six selected indicators appear directly in the open narrative field. These summaries use fixed local phrase patterns and store only counts and rates.

\begin{table}[p]
\centering
\caption{Aggregate narrative mentions for six selected CHAMPS indicators. Mention rate is the share of nonempty narratives matched by the pre-specified local phrase patterns. Precision is the structured-VA mean among observed structured responses with a mention. Sensitivity is the mention rate among structured-VA positives. The output stores counts and rates only; no narrative text, matched phrases, or row-level indicators are written.}
\label{tab:champs-goldilocks-mentions}
\small
\begin{tabularx}{\textwidth}{Xrrrrr}
\toprule
Indicator & Structured VA & Mention rate & Mentions & Precision & Sensitivity \\
\midrule
Doubts medical care was needed & 0.084 & 0.004 & 18 & 0.111 & 0.003 \\
Problems with treatment or respect & 0.150 & 0.010 & 45 & 0.368 & 0.021 \\
More than 2 hours to nearest facility & 0.148 & 0.019 & 88 & 0.164 & 0.021 \\
Traditional medicine used & 0.080 & 0.027 & 128 & 0.767 & 0.279 \\
Care costs displaced household payments & 0.242 & 0.044 & 206 & 0.208 & 0.039 \\
Convulsions or fits & 0.160 & 0.066 & 312 & 0.634 & 0.403 \\
\bottomrule
\end{tabularx}

\end{table}

\subsection{CHAMPS construct dictionary and data-handling rules}

The CHAMPS construct dictionary contains 30 binary constructs, a compact subset of the full verbal-autopsy item space. The goal is to evaluate AMV on analysis-level symptom, timing, death-location, care-seeking, access, and neonatal/perinatal quantities. Table~\ref{tab:champs-dictionary} reports the observed structured-verbal-autopsy denominator and observed mean for each construct. Denominators differ because some verbal-autopsy items are skipped, missing, or not applicable.

The source files used for this appendix and Section~\ref{sec:champs} are the restricted CHAMPS L2 files for verbal autopsy, basic demographics, and Determination of Cause of Death results from the 2026-06-01 export. Source records were treated as read-only. Restricted-data handling procedures prevent modification of source files and limit derived outputs to aggregate tables, figures, and manifests with suppression for small counts. Derived outputs exclude narrative text, case identifiers, vocabularies, embeddings, row-level mapped values, prompts, completions, snippets, top words, and model objects.

\begin{table}[p]
\centering
\caption{CHAMPS structured construct dictionary coverage. Observed \(n\) is the number of verbal-autopsy records with a nonmissing structured response for the derived construct. Mean yes is computed among observed structured responses. These structured responses define the structured-response scale for this analysis; they are not treated as clinical truth.}
\label{tab:champs-dictionary}
\footnotesize
\begin{tabularx}{\textwidth}{Xlrr}
\toprule
Construct & Domain & Observed $n$ & Mean yes \\
\midrule
Care costs displaced household payments & Access & 5,548 & 0.238 \\
Doubts medical care was needed & Access & 5,536 & 0.081 \\
More than 2 hours to nearest facility & Access & 5,579 & 0.135 \\
Care sought outside home & Care Pathway & 4,457 & 0.515 \\
Motorized transport to facility & Care Pathway & 3,803 & 0.854 \\
Phone or mobile call for help & Care Pathway & 5,888 & 0.306 \\
Problems during facility admission & Care Pathway & 3,941 & 0.112 \\
Problems getting medication or tests & Care Pathway & 3,926 & 0.102 \\
Problems with treatment or respect & Care Pathway & 3,936 & 0.130 \\
Traditional medicine used & Care Pathway & 5,582 & 0.087 \\
Traveled to health facility & Care Pathway & 5,581 & 0.682 \\
Treatment received during illness & Care Pathway & 4,871 & 0.752 \\
Death occurred in health facility & Death Location & 9,286 & 0.872 \\
Injury or accident & External & 6,690 & 0.019 \\
Assistance to breathe at birth & Neonatal Perinatal & 2,745 & 0.669 \\
Late-pregnancy complications & Neonatal Perinatal & 6,683 & 0.291 \\
Neonatal breathing problem & Neonatal Perinatal & 2,736 & 0.659 \\
Smaller than usual at birth & Neonatal Perinatal & 2,277 & 0.825 \\
Convulsions or fits & Symptoms & 4,951 & 0.138 \\
Cough & Symptoms & 4,984 & 0.131 \\
Difficulty breathing & Symptoms & 4,931 & 0.619 \\
Fever/infectious symptoms & Symptoms & 4,760 & 0.332 \\
Frequent loose or liquid stools & Symptoms & 4,998 & 0.143 \\
Gastrointestinal/diarrheal symptoms & Symptoms & 5,071 & 0.279 \\
Neurologic symptoms & Symptoms & 5,023 & 0.260 \\
Respiratory symptoms & Symptoms & 5,112 & 0.673 \\
Skin rash & Symptoms & 5,063 & 0.027 \\
Vomiting & Symptoms & 5,033 & 0.208 \\
Illness duration at least 3 days & Timing & 4,537 & 0.404 \\
Illness duration at least 7 days & Timing & 4,537 & 0.207 \\
\bottomrule
\end{tabularx}

\end{table}

\subsection{CHAMPS Validation-Item Support}

Table~\ref{tab:champs-design-support} records aggregate support values for a broader CHAMPS question-selection grid. This grid is separate from the design used for the displayed regressions in Section~\ref{sec:champs}. The displayed regressions use the sparse rule described in the main text: individual item reveals from the 30-construct dictionary and same-respondent regression-specific validation questions for about 9 percent of nonempty-narrative records. Appendix Table~\ref{tab:champs-regression-support} reports the realized support for those displayed regressions.

\begin{table}[p]
\centering
\caption{CHAMPS broader validation-item support grid. Mean items is the average number of structured validation items selected. Min item \(\pi\) is the smallest positive item probability. Same-respondent support is the share of records for which the full planned-analysis validation set is observed for the same respondent. This grid is separate from the sparse design used for the displayed regressions. Dashes indicate values suppressed in the saved aggregate output.}
\label{tab:champs-design-support}
\small
\begin{tabular}{lrrrr}
\toprule
Rule & $B$ & Mean items & Min item $\pi$ & Same-respondent support \\
\midrule
Regression set + random & 4 & 3.4 & 0.13 & 0.00 \\
Regression set + random & 8 & -- & -- & -- \\
Regression set + random & 12 & 8.6 & 0.40 & 0.01 \\
Regression set + random & 20 & 12.9 & 0.67 & 0.08 \\
Equal random & 4 & 3.4 & 0.13 & 0.00 \\
Equal random & 8 & -- & -- & -- \\
Equal random & 12 & 8.6 & 0.40 & 0.00 \\
Equal random & 20 & 12.9 & 0.67 & 0.07 \\
Constant score + regression set & 4 & 3.4 & 0.13 & 0.00 \\
Constant score + regression set & 8 & -- & -- & -- \\
Constant score + regression set & 12 & 8.6 & 0.40 & 0.00 \\
Constant score + regression set & 20 & 12.9 & 0.67 & 0.07 \\
\bottomrule
\end{tabular}

\end{table}

\section{Disclosure Elements}
\label{app:disclosure}

This article reports a statistical design and three empirical illustrations. It includes no new respondent-facing data collection. The empirical sections use a design-calibration simulation, public ATUS files, and a restricted de-identified CHAMPS L2 export analyzed locally. Table~\ref{tab:disclosure-elements} summarizes the disclosure elements most relevant for initial review.

\begin{table}[p]
\centering
\caption{Disclosure elements for empirical materials}
\label{tab:disclosure-elements}
\footnotesize
\begin{tabularx}{\textwidth}{p{0.18\textwidth}X}
\toprule
Element & Disclosure \\
\midrule
New human-subject data collection & None. The paper proposes a future survey design and evaluates it through simulations and analyses using existing survey records. No respondent-facing AI interview was fielded for this article. \\
Design-calibration simulation & Section~\ref{sec:calibration} uses simulated finite populations with \(n=5{,}000\), 800 Monte Carlo repetitions, and validation probabilities \(q=0.05,0.10,0.25\). The simulation examines estimator behavior in a controlled setting and does not report a survey estimate. \\
ATUS data & Section~\ref{sec:atus} uses public ATUS respondent and activity files for 2018, 2019, and 2021--2024. The analytic file contains 52,468 respondent-days and 970,712 activity episodes. ATUS final weights are used as supplied by BLS in a pooled respondent-day estimand over included years. The simulation standard errors reported in Tables~\ref{tab:atus-items} and \ref{tab:atus-regressions} are linearized approximations used to compare methods within the simulation, not full ATUS replicate-weight sampling variances. \\
CHAMPS data & Section~\ref{sec:champs} uses the de-identified CHAMPS L2 export dated 2026-06-01. The verbal-autopsy file contains 9,299 records; 4,693 have nonempty designated narrative fields and 4,606 are blank or missing. No design weights were identified or used, so CHAMPS results are unweighted summaries of analyzable VA records in the selected export. Displayed item estimates are conditional on nonempty narratives and observed structured responses for the construct. \\
Question wording and structured response variables & The paper introduces no fielded questionnaire. ATUS structured response variables are derived from public diary activity records. CHAMPS structured response variables are analysis-level constructs derived from structured VA fields and listed in Table~\ref{tab:champs-dictionary}. A future field implementation would need to include the embedded validation-question wording and versioned protocol. \\
Privacy and restricted data & CHAMPS source files were read in place and not modified. Derived CHAMPS outputs are aggregate-only with \(n<10\) suppression. The analysis writes no raw narratives, case identifiers, vocabularies, embeddings, model objects, prompts, completions, snippets, top words, or row-level mapped values. \\
Replication materials & Replication materials include code, configuration files, analysis manifests, seeds, and aggregate outputs for the design-calibration, ATUS, and CHAMPS analyses. ATUS uses public source data. Reproducing CHAMPS tables requires authorized access to the restricted CHAMPS L2 export; public replication materials should include aggregate outputs and code but not restricted source records or raw narrative text. \\
\bottomrule
\end{tabularx}
\end{table}

\end{document}